\documentclass[a4paper,fleqn]{cas-sc}

\usepackage[authoryear]{natbib}
\usepackage{bm}
\usepackage{mathtools}
\usepackage{graphicx}
\usepackage{url}
\usepackage[capitalise]{cleveref}

\def\tsc#1{\csdef{#1}{\textsc{\lowercase{#1}}\xspace}}
\tsc{WGM}
\tsc{QE}

\begin{document}
\let\WriteBookmarks\relax
\def\floatpagepagefraction{1}
\def\textpagefraction{.001}

\shorttitle{Reduced order micromorphic homogenization of metamaterials}

\shortauthors{Guo and Rokoš}

\title [mode = title]{Reduced order modeling for micromorphic computational homogenization of mechanical metamaterials with patterning fluctuation fields}



%

\author[1]{Theron Guo}[orcid=0000-0002-8465-273X]

\cormark[1]

\ead{t.guo@tue.nl}

\credit{Conceptualization, Methodology, Software, Validation, Investigation, Data curation, Visualization, Writing -- original draft}

\affiliation[1]{organization={Mechanics of Materials, Eindhoven University of Technology},
            country={The Netherlands}}

\author[1]{Ond\v{r}ej Roko\v{s}}[orcid=0000-0003-2589-5333]

\ead{o.rokos@tue.nl}

\credit{Conceptualization, Methodology, Writing -- review \& editing}

\cortext[1]{Corresponding author}


\begin{abstract}
Mechanical metamaterials owe some of their exotic effective properties to microstructural instabilities such as buckling and pattern transformation. Their fine-scale geometry, combined with the interplay of local and nonlocal effects, renders direct numerical simulation (DNS) prohibitively expensive, if not intractable. Micromorphic computational homogenization~\citep{Rokos2019} has been proposed to make macroscopic simulations feasible: a representative volume element (RVE) carrying patterning fluctuation fields resolves the microscale, while an effective micromorphic continuum describes the macroscale. Such simulations nevertheless remain costly, since the RVE problem and its associated tangent problems must be solved at every macroscopic integration point, which becomes prohibitive, especially in three dimensions. We therefore propose a reduced order model for the microscopic problem, combining proper orthogonal decomposition with a hyperreduction method tailored to the micromorphic setting. Two numerical examples quantify the reduced basis and hyperreduction errors and discuss the offline costs and attainable online speed-ups. The results show that this approach can enable engineering-scale large deformation analysis of mechanical metamaterials.
\end{abstract}


\begin{highlights}
\item A novel reduced order model for the RVE problem of micromorphic homogenization.
\item Dedicated reduced integration rule for integrating effective stress and mode-conjugate terms.
\item Reduced basis and hyperreduction errors quantified for two examples.
\item Two-scale buckling analysis of a 3D lattice metamaterial.
\end{highlights}


\begin{keywords}
Computational homogenization \sep Hyperreduction \sep Mechanical metamaterials \sep Bifurcation \sep Large deformation
\end{keywords}

\maketitle

\section{Introduction}\label{sec:intro}

Metamaterials do not derive their macroscopic behavior from the constituents of which they are made, but from the shape and topology of the microstructure. Architecting this microstructure gives access to effective properties that are difficult or impossible to realize with conventional solids. Prominent examples are a negative Poisson's ratio, obtained from re-entrant or perforated architectures~\citep{Bertoldi2010,Babaee2013}; acoustic band gaps whose location and width follow from the periodic arrangement and can subsequently be modulated by deformation~\citep{Kushwaha1993,BertoldiBoyce2008}; or thermal cloaks that steer the flow of heat around a protected region~\citep{Guenneau2012,Narayana2012,Schittny2013}. Broad overviews are given, e.g., by \citet{Bertoldi2017} and \citet{Kadic2019}.

Among these, \emph{mechanical metamaterials} made of elastomers or of slender lattices are of particular interest, since they can be designed to operate deep in the large deformation regime and exploit elastic instabilities as a design mechanism rather than avoiding them as a failure mode: beyond a critical load, the microstructure buckles into a reversible, spatially correlated pattern, which abruptly alters the effective stiffness, the Poisson's ratio or the band structure~\citep{Mullin2007,Bertoldi2008}. This makes them attractive, e.g., for soft robotics and biomedical applications~\citep{Rafsanjani2019,Kolken2017}, where shape morphing and large deformations are necessary. Experiments have demonstrated programmable and combinatorial designs whose response can be reprogrammed through the boundary conditions or the loading history~\citep{Florijn2014,Coulais2016,Coulais2018b}, anomalous size effects traced back to an intrinsic length scale of the buckling pattern~\citep{Coulais2018}, and three-dimensional architectures with unconventional couplings such as twist~\citep{Frenzel2017}.

Exploiting this design freedom requires simulation tools that link the microstructural geometry to the macroscopic response. In principle, a direct numerical simulation (DNS) resolving every geometric feature of the specimen provides exactly this link, and it remains the reference against which any model has to be judged. In practice, its cost grows prohibitive as soon as the microstructure becomes fine relative to the specimen, all the more so when the instabilities of interest have to be detected and the solution has to be continued beyond them. In three dimensions, DNS is therefore typically restricted to small specimens or to beam and truss representations of the lattice~\citep{Radi2024}, which quasicontinuum methods extend towards larger scales by resolving the lattice only where needed~\citep{Phlipot2019,Kraschewski2024}. Multiscale methods trade this resolution for a homogenized macroscopic description in which the microstructure is represented only through a smaller auxiliary problem defined on a representative volume element (RVE). 

In the small deformation regime, such tools are well established. Beam and truss lattice models reduce the slender members of an architected material to one-dimensional elements and thereby render simulations of an entire specimen affordable~\citep{Vigliotti2012}, and can be homogenized further into effective second-gradient continua~\citep{Weeger2021}, while classical, first-order computational homogenization (CH1) --- in which an RVE is solved at every macroscopic integration point --- delivers effective properties for arbitrary microstructures, under the assumption that the microscopic length scale is much smaller than the macroscopic one; see~\citet{Geers2010} and \citet{Matous2017} for comprehensive overviews, and~\citep{Smit1998,Miehe1999,Feyel2000} for the underlying two-scale (FE\textsuperscript{2}) framework. At large deformations the situation is less settled. Most existing work presupposes that the CH1 conditions hold, in which case a single RVE driven by the prescribed macroscopic deformation gradient is sufficient to determine the effective properties; this is the route taken for hyperelastic beams and beam networks~\citep{LeClezio2023,LeClezio2024,Gaertner2021} and for periodic truss lattices at finite strains~\citep{Glaesener2019,Glaesener2020}. These conditions require the macroscopic fields to vary slowly over the size of the RVE, and it is precisely this requirement that microstructural buckling violates: the instability introduces a length scale of its own, the effective energy loses rank-one convexity, and the response softens over a region whose extent is dictated by the microstructure rather than by the macroscopic loading~\citep{Geymonat1993,Triantafyllidis1996}. First-order predictions then deteriorate and become mesh-dependent, and the associated error has been shown to grow rapidly once scale separation becomes marginal~\citep{Ameen2018,Ameen2018b,Sperling2024}.

For this reason, several enriched homogenization schemes have been developed, which equip the macroscopic continuum with additional kinematic information and, with it, an intrinsic length scale. In second-order computational homogenization (CH2), the macroscopic deformation gradient is supplemented by its gradient, so that the RVE is driven by a quadratic rather than a linear macroscopic field and the macrostructure becomes a second-gradient continuum~\citep{Kouznetsova2002,Kouznetsova2004}. A related enrichment was proposed by \citet{Biswas2017} and \citet{Biswas2020}, in which an additional macroscopic field describing the average strain of the microscopic constituents is transmitted between the scales. For buckling and pattern-transforming metamaterials specifically, both CH1 and CH2 run into issues caused by the underlying bifurcation: CH1 becomes mesh-dependent, as a direct consequence of the loss of rank-one convexity~\citep{Geymonat1993,Triantafyllidis1996}, and suffers from convergence issues where the response localizes, whereas CH2 fails to accurately capture the bifurcation strain and introduces an overly stiff pre-buckling response~\citep{Sperling2024}. This motivated the micromorphic scheme of \citet{Rokos2019}, in which the microscopic displacement is enriched with a small set of precomputed patterning modes whose amplitudes become independent macroscopic fields, so that the buckling pattern is represented explicitly and its spatial variation is governed by its own balance equation. These enriched schemes have, however, been developed and demonstrated almost exclusively in two dimensions, and the few three-dimensional demonstrations remain restricted to small strains; we are unaware of any three-dimensional study at large deformations. The reason is cost: each enrichment adds macroscopic degrees of freedom and, more importantly, further work-conjugate quantities and tangent blocks, every one of which requires an additional sensitivity solve on an RVE that must itself be resolved in three dimensions --- at every macroscopic integration point, in every macroscopic Newton iteration, and in every load step.

A comparable bottleneck, in its milder classical form, has long been recognized for CH1, where considerable effort has gone into replacing the microscopic problem with a cheap-to-evaluate surrogate, along two broad strategies. The first is projection-based: the microscopic fluctuation field is approximated in a low-dimensional basis, typically obtained through proper orthogonal decomposition (POD)~\citep{Quarteroni2016,Hesthaven2016}, and the residual dependence on the full microscopic mesh is subsequently removed by hyperreduction, for which the empirical interpolation method~\citep{Barrault2004} and its discrete counterpart~\citep{Chaturantabut2010}, energy-conserving sampling and weighting~\citep{Farhat2014}, and the empirical cubature method (ECM)~\citep{Hernandez2017} are the established choices. Reduced order models (ROMs) of this type have been applied to hyperelastic and elasto-plastic RVEs~\citep{Yvonnet2007,Fritzen2013,Hernandez2014} and, in our own previous works~\citep{Guo2021,Guo2022,Guo2024reduced}. The second strategy dispenses with the microscopic problem altogether and learns an effective constitutive model directly from RVE data, using neural networks that are either generic~\citep{Le2015,Bessa2017} or constrained to respect thermodynamic and material-symmetry requirements~\citep{Linka2021,Masi2021,Kalina2023,Gaertner2021,Ghane2026}. The two differ markedly in their data requirements. Because a projection-based ROM inherits the governing equations and needs only enough snapshots to span the solution manifold, it is typically very data efficient, whereas a learned constitutive model must sample the entire macroscopic input space and generally requires substantially more data --- although embedding physical constraints reduces this demand considerably. In either case, the surrogate is only as good as the homogenization scheme it accelerates, and for the softening, pattern-forming response of interest here the underlying CH1 assumptions themselves no longer hold.

For the enriched formulations, surrogate modeling has received far less attention, and the balance between the two strategies shifts. Learning an effective constitutive model becomes considerably less attractive, because the input space is no longer the macroscopic deformation gradient alone: CH2 additionally requires its full gradient, and the micromorphic scheme the mode amplitudes together with their gradients. Meaningful sampling bounds for these gradient quantities are difficult to establish a priori, while every training sample is itself a solve of the enriched microscopic problem. Projection-based reduction, being far more data efficient, is therefore the more natural candidate, as we demonstrated for CH2 in two dimensions~\citep{Guo2025}, where POD combined with a tailored, ECM-inspired hyperreduction reproduced the full second-order response at speed-ups on the order of 100 relative to a DNS. Recently, \citet{Ju2026} followed a similar POD-based route for the scheme of \citet{Biswas2017}, applied to elasto-plastic porous materials.

To the best of our knowledge, no ROM has so far been proposed for the micromorphic scheme of \citet{Rokos2019}, which is the enrichment best suited to the pattern-transforming metamaterials considered here. In this work, we address this gap. Our main contributions are:
\begin{enumerate}
\item a POD- and hyperreduction-based ROM of the micromorphic microscopic problem, together with a consistent derivation of the effective stress, of the quantities conjugate to the mode amplitudes and their gradients, and of their derivatives with respect to all macroscopic driving quantities, which together form the macroscopic tangent;
\item a two-dimensional numerical study, in which reference solutions are still affordable, validating the implementation against a DNS and quantifying the reduced basis and the hyperreduction errors separately;
\item a fully three-dimensional two-scale buckling analysis of a lattice metamaterial, for which neither a DNS nor the full micromorphic model is computationally feasible, as an outlook on the class of problems the approach brings within reach;
\item and a discussion of the offline costs and attainable online speed-ups.
\end{enumerate}

The remainder of this paper is organized as follows. In \cref{sec:formulation}, the micromorphic computational homogenization framework is summarized, together with the microscopic problem and the effective quantities it has to deliver. \Cref{sec:rom} develops the reduced order model, covering the POD basis and the hyperreduction of the micromorphic-specific quantities. In \cref{sec:examples}, the proposed method is examined in detail, first for the two-dimensional example and subsequently for the three-dimensional one. Finally, \cref{sec:conclusions} summarizes the findings and closes with some concluding remarks.

\subsection{Notation}\label{sec:intro:notation}

Throughout the article, the following notation conventions are used:
\begin{itemize}
\item scalars $a$,
\item vectors $\bm{a} = a_i \bm{e}_i$,
\item position vectors $\bar{\bm{X}}$ (macroscopic body) and $\bm{X}$ (microscopic representative volume element), both taken in the reference configuration,
\item second-order tensors $\bm{A} = A_{ij}\, \bm{e}_i \bm{e}_j$,
\item third- and fourth-order tensors $\mathcal{T} = \mathcal{T}_{ijk}\, \bm{e}_i \bm{e}_j \bm{e}_k$ and $\mathbb{A} = \mathbb{A}_{ijkl}\, \bm{e}_i \bm{e}_j \bm{e}_k \bm{e}_l$,
\item matrices $\mathbf{A}$ and column matrices $\mathbf{a}$,
\item $\bm{a}\cdot\bm{b} = a_ib_i$, $\ \bm{a}\otimes\bm{b} = a_ib_j\,\bm{e}_i\bm{e}_j$, $\ \bm{A}\cdot\bm{b} = A_{ij}b_j\,\bm{e}_i$, $\ \bm{A}\cdot\bm{B} = A_{ik}B_{kj}\,\bm{e}_i\bm{e}_j$, $\ \bm{A}:\bm{B} = A_{ij}B_{ij}$,
\item contractions of higher-order tensors, $\mathbb{A}:\bm{B} = \mathbb{A}_{ijkl}B_{kl}\,\bm{e}_i\bm{e}_j$ for a fourth-order tensor, and $\mathcal{T}:\bm{B} = \mathcal{T}_{ijk}B_{jk}\,\bm{e}_i$, $\ \mathcal{T}\cdot\bm{c} = \mathcal{T}_{ijk}c_k\,\bm{e}_i\bm{e}_j$ for a third-order tensor,
\item gradient and divergence operators with respect to the reference configuration, $\nabla a = \tfrac{\partial a}{\partial X_i}\bm{e}_i$ for a scalar $a$, $\ \nabla\bm{a} = \tfrac{\partial a_j}{\partial X_i}\bm{e}_i\bm{e}_j$ for a vector $\bm{a}$, and $\ \nabla\cdot\bm{A} = \tfrac{\partial A_{ij}}{\partial X_i}\bm{e}_j$, here written for the microscale and defined analogously with respect to $\bar{\bm{X}}$ at the macroscale,
\item volume average $\langle \bullet \rangle \coloneqq \tfrac{1}{|\Omega|}\int_\Omega \bullet\, \mathrm{d}\bm{X}$ over the microscopic representative volume element $\Omega$ with volume $|\Omega|$,
\item Euclidean norm $\|\bm{a}\| \coloneqq \sqrt{\bm{a}\cdot\bm{a}}$ of a vector, and analogously $\|\mathbf{a}\| \coloneqq \sqrt{\mathbf{a}^T\mathbf{a}}$ of a column matrix,
\item linearization of a functional $\mathcal{F}$ around a state $\bm{a}$ in direction $\Delta\bm{a}$, $\ D\mathcal{F}|_{\bm{a}}\cdot(\Delta\bm{a}) \coloneqq \tfrac{d}{d\tau}\mathcal{F}(\bm{a}+\tau\Delta\bm{a})\big|_{\tau=0}$, with $D^2\mathcal{F}\cdot[\Delta\bm{a},\delta\bm{a}]$ the corresponding second variation, the base state being omitted wherever it is clear from the context,
\end{itemize}
where the Einstein summation convention is assumed on repeated indices $i,j,k,l = 1,\dots,d$, with $d$ the spatial dimension, and $\bm{e}_i$ the basis vectors of a $d$-dimensional Cartesian coordinate frame. The index labeling the enrichment modes introduced below in \cref{sec:formulation} is also written $i$, but is never subject to the summation convention; wherever it is summed, an explicit summation sign is used. An overline marks macroscopic quantities, as in $\bar{\bm{X}}$, $\bar{\bm{u}}$ and $\bar{\bm{F}}$ versus their microscopic counterparts $\bm{X}$, $\bm{u}$ and $\bm{F}$. This applies also to macroscopic quantities without a microscopic namesake, such as the rotation $\bar{\bm{R}}$ from the polar decomposition of $\bar{\bm{F}}$, the enrichment amplitudes $\bar{v}_i$ and their gradients $\bar{\bm{g}}_i$, and the work-conjugate quantities $\bar{\Gamma}_i$ and $\bar{\bm{\Lambda}}_i$, all introduced below in \cref{sec:formulation}.

\section{Problem formulation}\label{sec:formulation}

\subsection{Macroscopic problem}\label{sec:formulation:macro}

We follow the micromorphic computational homogenization formulation of \citet{Rokos2019}, which we briefly summarize here; the reader is referred to~\citet{Rokos2019} and \citet{vanBree2020} for a more detailed exposition. Consider a macroscopic body $\mathcal{B} \subset \mathbb{R}^d$, $d = 2,3$, with boundary $\partial\mathcal{B}$ and reference position vector $\bar{\bm{X}} \in \mathcal{B}$. Unlike the classical (first-order) macroscopic continuum, the micromorphic continuum is parameterized not only by the macroscopic displacement field $\bar{\bm{u}}(\bar{\bm{X}})$, but also by $N_\phi$ scalar enrichment-amplitude fields $\bar{v}_i(\bar{\bm{X}})$, $i = 1,\dots,N_\phi$, which carry the macroscopic imprint of the microstructural patterning modes $\bm{\phi}_i$ introduced below in \cref{sec:formulation:micro}. The macroscopic deformation gradient and the spatial gradient of the amplitude fields are given by
\begin{equation}
\bar{\bm{F}} \coloneqq \bm{I} + (\nabla\bar{\bm{u}})^T, \qquad \bar{\bm{g}}_i \coloneqq \nabla \bar{v}_i,
\label{eq:macro_kin}
\end{equation}
which, together with $\bar{v}_i$, act as the macroscopic driving quantities of the microscopic problem, and enter the macroscopic energy density $\bar{\Psi}(\bar{\bm{F}}, \bar{v}_i, \bar{\bm{g}}_i)$, obtained from homogenization of the microscopic strain energy over the RVE, as detailed below in \cref{sec:formulation:effective}. The total macroscopic energy then reads
\begin{equation}
\mathcal{E}[\bar{\bm{u}}, \{\bar{v}_i\}] \coloneqq \int_{\mathcal{B}} \bar{\Psi}(\bar{\bm{F}}, \bar{v}_i, \bar{\bm{g}}_i)\, \mathrm{d}\bar{\bm{X}}.
\label{eq:macro_energy}
\end{equation}
The first variation of \cref{eq:macro_energy} with respect to $(\bar{\bm{u}}, \{\bar{v}_i\})$, in the directions $(\delta\bar{\bm{u}}, \{\delta \bar{v}_i\})$, yields the coupled weak form
\begin{equation}
D\mathcal{E}\cdot(\delta\bar{\bm{u}}, \{\delta \bar{v}_i\}) = \int_{\mathcal{B}} (\nabla\delta\bar{\bm{u}})^T : \bar{\bm{P}}\, \mathrm{d}\bar{\bm{X}} + \sum_{i=1}^{N_\phi}\int_{\mathcal{B}} \big(\bar{\Gamma}_i\, \delta \bar{v}_i + \bar{\bm{\Lambda}}_i \cdot \nabla\delta \bar{v}_i\big)\, \mathrm{d}\bar{\bm{X}} \overset{!}{=} 0,
\label{eq:macro_weak}
\end{equation}
for all admissible test functions $(\delta\bar{\bm{u}}, \{\delta \bar{v}_i\})$, where $\bar{\bm{P}} \coloneqq \partial\bar{\Psi}/\partial\bar{\bm{F}}$, $\bar{\Gamma}_i \coloneqq \partial\bar{\Psi}/\partial \bar{v}_i$ and $\bar{\bm{\Lambda}}_i \coloneqq \partial\bar{\Psi}/\partial\bar{\bm{g}}_i$ are, respectively, the effective stress and the mode-conjugate scalar and vector quantities, whose evaluation from the microscopic solution is detailed below in \cref{sec:formulation:effective}. The problem is closed by Dirichlet boundary conditions on both fields,
\begin{equation}
\bar{\bm{u}} = \hat{\bar{\bm{u}}} \quad \text{on } \partial\mathcal{B}_{\bar{\bm{u}}}, \qquad \bar{v}_i = \hat{\bar{v}}_i \quad \text{on } \partial\mathcal{B}_{\bar{v}_i}, \quad i = 1,\dots,N_\phi,
\label{eq:macro_bc}
\end{equation}
prescribed on (possibly different) parts $\partial\mathcal{B}_{\bar{\bm{u}}}, \partial\mathcal{B}_{\bar{v}_i} \subseteq \partial\mathcal{B}$ of the boundary, on which the test functions $\delta\bar{\bm{u}}$ and $\delta \bar{v}_i$ vanish accordingly. On the remainder of the boundary, the corresponding natural boundary conditions,
\begin{equation}
\bar{\bm{P}}\cdot\bar{\bm{N}} = \bm{0} \quad \text{on } \partial\mathcal{B}\setminus\partial\mathcal{B}_{\bar{\bm{u}}}, \qquad \bar{\bm{\Lambda}}_i\cdot\bar{\bm{N}} = 0 \quad \text{on } \partial\mathcal{B}\setminus\partial\mathcal{B}_{\bar{v}_i},
\label{eq:macro_bc_nat}
\end{equation}
follow from \cref{eq:macro_weak}, with $\bar{\bm{N}}$ the outward unit normal of $\partial\mathcal{B}$ in the reference configuration.

Since \cref{eq:macro_weak} is nonlinear in $(\bar{\bm{u}}, \{\bar{v}_i\})$, it is solved with Newton's method, which requires the second variation of \cref{eq:macro_energy},
\begin{equation}
\begin{split}
D^2\mathcal{E}\cdot\big[(\Delta\bar{\bm{u}}, \{\Delta \bar{v}_i\}), (\delta\bar{\bm{u}}, \{\delta \bar{v}_i\})\big] &= \int_{\mathcal{B}} \bigg[ (\nabla\delta\bar{\bm{u}})^T : \frac{\partial\bar{\bm{P}}}{\partial\bar{\bm{F}}} : (\nabla\Delta\bar{\bm{u}})^T \\
&\quad + \sum_{i=1}^{N_\phi} \Big((\nabla\delta\bar{\bm{u}})^T : \frac{\partial\bar{\bm{P}}}{\partial \bar{v}_i}\, \Delta \bar{v}_i + (\nabla\delta\bar{\bm{u}})^T : \frac{\partial\bar{\bm{P}}}{\partial\bar{\bm{g}}_i}\cdot\nabla\Delta \bar{v}_i\Big) \bigg]\, \mathrm{d}\bar{\bm{X}} \;+\; \dots,
\end{split}
\label{eq:macro_lin}
\end{equation}
where the omitted terms follow analogously from the derivatives of $\bar{\Gamma}_i$ and $\bar{\bm{\Lambda}}_i$; the complete expression is given in Appendix~\ref{app:linearization}. Together with those of $\bar{\bm{P}}$ shown above, these derivatives with respect to $\bar{\bm{F}}$, $\bar{v}_i$ and $\bar{\bm{g}}_i$ form a $3\times3$ block structure of macroscopic tangents --- the three work-conjugate quantities $(\bar{\bm{P}}, \bar{\Gamma}_i, \bar{\bm{\Lambda}}_i)$ differentiated with respect to the three macroscopic driving quantities $(\bar{\bm{F}}, \bar{v}_i, \bar{\bm{g}}_i)$ --- each block of which is computed from the microscopic sensitivity problem of \cref{sec:formulation:effective} below. Given a finite element discretization of $\bar{\bm{u}}$ and $\bar{v}_i$, \cref{eq:macro_weak,eq:macro_lin} are assembled into a global residual and tangent stiffness matrix, which are then driven to convergence by the Newton iteration.

\subsection{Microscopic problem}\label{sec:formulation:micro}

At the microscale, the classical periodic homogenization ansatz is enriched with a finite family of $N_\phi$ user-supplied global modes $\bm{\phi}_i(\bm{X})$, $i = 1,\dots,N_\phi$, defined on the reference configuration of an RVE $\Omega \subset \mathbb{R}^d$, together with macroscopic amplitude variables $\bar{v}_i \in \mathbb{R}$ and gradient variables $\bar{\bm{g}}_i \in \mathbb{R}^d$. For brevity, dependence of the microscopic quantities on the macroscopic position is omitted throughout, and a fixed macroscopic material point is implicitly assumed.

\paragraph{Displacement.}
The total microscopic displacement field $\bm{u}(\bm{X})$ is decomposed into a mean field, prescribed by the macroscopic deformation gradient $\bar{\bm{F}}$ and the enrichment modes, and a fluctuation field $\bm{w}(\bm{X})$,
\begin{equation}
\bm{u}(\bm{X}) = (\bar{\bm{F}} - \bm{I})\cdot\bm{X} + \sum_{i=1}^{N_\phi} \big(\bar{v}_i + \bm{X}\cdot\bar{\bm{g}}_i\big)\, \bar{\bm{R}}\cdot\bm{\phi}_i(\bm{X}) + \bm{w}(\bm{X}),
\label{eq:utot}
\end{equation}
where $\bm{I}$ is the second-order identity tensor and $\bar{\bm{R}}$ is the rotation tensor obtained from the polar decomposition $\bar{\bm{F}} = \bar{\bm{R}}\cdot\bar{\bm{U}}$ of the macroscopic deformation gradient into a rotation $\bar{\bm{R}}$ and a symmetric stretch $\bar{\bm{U}}$.

\textit{Remark.} \citet{Rokos2019} and \citet{vanBree2020} keep the modes $\bm{\phi}_i$ fixed in the reference frame, i.e., $\bar{\bm{R}} = \bm{I}$ in \cref{eq:utot}. Here, the modes are instead co-rotated with the macroscopic rotation $\bar{\bm{R}}$. This is essential for the homogenized constitutive response to remain frame-indifferent, since $\bm{\phi}_i$ are in general not isotropic and a fixed-mode ansatz would otherwise let a superposed rigid rotation of the RVE alter the effective stress. It has, in addition, a useful computational consequence, to which we return at the end of this section.

\paragraph{Deformation gradient.}
The microscopic deformation gradient follows from \cref{eq:utot} as
\begin{equation}
\bm{F} \coloneqq \bm{I} + (\nabla\bm{u})^T = \bar{\bm{F}} + \sum_{i=1}^{N_\phi}\Big[(\bar{\bm{R}}\cdot\bm{\phi}_i) \otimes \bar{\bm{g}}_i + \big(\bar{v}_i + \bm{X}\cdot\bar{\bm{g}}_i\big)\big(\nabla(\bar{\bm{R}}\cdot\bm{\phi}_i)\big)^T\Big] + (\nabla\bm{w})^T.
\label{eq:Fmicro}
\end{equation}
Note that $\bar{\bm{F}}$, $\bar{v}_i$ and $\bar{\bm{g}}_i$ enter \cref{eq:Fmicro} as prescribed macroscopic quantities, while $\bm{w}$ is the only remaining unknown of the microscopic problem.

\paragraph{Weak form.}
Neglecting body forces, the quasi-static microscopic equilibrium is expressed in weak form as
\begin{equation}
G(\bm{w}; \delta\bm{w}) \coloneqq \int_\Omega (\nabla\delta\bm{w})^T : \bm{P}(\bm{F})\, \mathrm{d}\bm{X} \overset{!}{=} 0,
\label{eq:weak}
\end{equation}
for all admissible test functions $\delta\bm{w}$, where $\bm{P}$ is the microscopic first Piola--Kirchhoff stress tensor, generally a nonlinear function of $\bm{F}$ through a given microscopic constitutive law, and where $\bar{\bm{F}}$, $\bar{v}_i$ and $\bar{\bm{g}}_i$ act as external forcing terms entering through \cref{eq:Fmicro}. No constitutive model is specified at this stage; the microscopic problem is stated here for a generic hyperelastic material.

\paragraph{Linearization.}
Since \cref{eq:weak} is generally nonlinear in $\bm{w}$, it is solved with Newton's method, which requires the linearization of \cref{eq:weak} with respect to $\bm{w}$ in the direction $\Delta\bm{w}$,
\begin{equation}
DG|_{\bm{w}}\cdot(\Delta\bm{w}) = \int_\Omega (\nabla\delta\bm{w})^T : \mathbb{A}(\bm{F}) : (\nabla\Delta\bm{w})^T\, \mathrm{d}\bm{X},
\label{eq:lin}
\end{equation}
with $\mathbb{A} \coloneqq \partial\bm{P}/\partial\bm{F}$ the microscopic tangent stiffness tensor.

\paragraph{Constraints.}
The decomposition in \cref{eq:utot} is unique only if periodic boundary conditions are imposed on $\bm{w}$, together with three families of constraints that prevent $\bm{w}$ from re-absorbing rigid-body motion or any contribution already carried by the enrichment modes~\citep{vanBree2020}:
\begin{align}
\langle \bm{w} \rangle &= \bm{0}, \label{eq:c1}\\
\langle \bm{w}\cdot\bar{\bm{R}}\cdot\bm{\phi}_i \rangle &= 0, \quad i = 1,\dots,N_\phi, \label{eq:c2}\\
\langle (\bm{w}\cdot\bar{\bm{R}}\cdot\bm{\phi}_i)\, \bm{X} \rangle &= \bm{0}, \quad i = 1,\dots,N_\phi, \label{eq:c3}
\end{align}
where $\langle\bullet\rangle$ denotes the volume average. \Cref{eq:c1} fixes the rigid-body translation, while \cref{eq:c2,eq:c3} enforce orthogonality of the fluctuation field to the patterning modes and their linear moments, so that $\bar{v}_i$ and $\bar{\bm{g}}_i$ remain uniquely identified as the amplitude and gradient of mode $i$, respectively. Note that the normalization of the modes is arbitrary: rescaling $\bm{\phi}_i \rightarrow \alpha\,\bm{\phi}_i$ is exactly compensated by $\bar{v}_i \rightarrow \bar{v}_i/\alpha$ and $\bar{\bm{g}}_i \rightarrow \bar{\bm{g}}_i/\alpha$ (with the work-conjugate quantities $\bar{\Gamma}_i$ and $\bar{\bm{\Lambda}}_i$ of \cref{sec:formulation:macro} scaling accordingly), leaving the total displacement of \cref{eq:utot} and all other effective quantities unchanged.

\paragraph{Discretization.}
The fluctuation field is discretized with finite elements following standard practice, see, e.g.,~\citep{Guo2025}. Periodicity and the constraints in \cref{eq:c1,eq:c2,eq:c3} are enforced through Lagrange multipliers $\mathbf{m}$, which turns the discretized problem into a saddle-point system. Given the current Newton iterate $\mathbf{w}_k$, the discrete update follows from
\begin{equation}
\begin{bmatrix} \mathbf{K}(\mathbf{w}_k) & \mathbf{C}^T \\ \mathbf{C} & \mathbf{0} \end{bmatrix}
\begin{bmatrix} \Delta\mathbf{w} \\ \mathbf{m} \end{bmatrix}
=
\begin{bmatrix} -\mathbf{f}(\mathbf{w}_k) \\ \mathbf{0} \end{bmatrix},
\qquad \mathbf{w}_{k+1} = \mathbf{w}_k + \Delta\mathbf{w},
\label{eq:newton}
\end{equation}
where $\mathbf{K}$ is the global tangent stiffness matrix assembled from \cref{eq:lin}, $\mathbf{f}$ is the residual assembled from \cref{eq:weak}, and $\mathbf{C}$ collects the discretized periodicity and constraint equations. \Cref{eq:newton} is iterated to convergence, i.e., until $\|\mathbf{f}(\mathbf{w}_k) + \mathbf{C}^T\mathbf{m}\| \leq \varepsilon_{\mathrm{newton}}$ for a user-defined tolerance $\varepsilon_{\mathrm{newton}}$. In our implementation the multipliers are not formed explicitly: the periodicity ties are eliminated directly at the level of the degrees of freedom, and the constraints of \cref{eq:c1,eq:c2,eq:c3} by projecting \cref{eq:newton} onto their null space. Both eliminations are equivalent to the bordered system written above, but yield a smaller and symmetric system.

\paragraph{Solution in the material frame.}
We can now return to the computational consequence of the co-rotated ansatz. Because the modes rotate with the macrostructure, the RVE problem never needs to be solved for a rotated state: substituting $\bar{\bm{F}} = \bar{\bm{R}}\cdot\bar{\bm{U}}$ into \cref{eq:utot} shows that the solution for $(\bar{\bm{F}}, \bar{v}_i, \bar{\bm{g}}_i)$ is the rigid rotation by $\bar{\bm{R}}$ of the solution for $(\bar{\bm{U}}, \bar{v}_i, \bar{\bm{g}}_i)$. In practice the RVE is therefore always driven by the symmetric stretch $\bar{\bm{U}}$, i.e., \cref{eq:utot,eq:Fmicro} are evaluated with $\bar{\bm{R}} = \bm{I}$ and $\bar{\bm{F}}$ replaced by $\bar{\bm{U}}$, and the resulting effective quantities are rotated back afterwards, as detailed below in \cref{sec:formulation:effective}. We refer to this setting as the \emph{material frame} in what follows, as opposed to the \emph{lab frame}, in which the macroscopic problem is posed and in which the RVE is subjected to the full $\bar{\bm{F}}$, rotation $\bar{\bm{R}}$ included. Since $\bar{\bm{U}}$ is symmetric, it also reduces the number of macroscopic deformation directions --- both those sampled when generating training data and those for which a tangent has to be computed --- from $d^2$ to $d(d+1)/2$, i.e., from 4 to 3 in two dimensions or from 9 to 6 in three dimensions.

\subsection{Effective quantities}\label{sec:formulation:effective}

After the microscopic problem has been solved and a converged fluctuation field $\bm{w}^*$ obtained, the effective stress $\bar{\bm{P}}$, the mode-conjugate scalars $\bar{\Gamma}_i$ and the mode-conjugate vectors $\bar{\bm{\Lambda}}_i$, together with their derivatives with respect to $\bar{\bm{F}}$, $\bar{v}_i$ and $\bar{\bm{g}}_i$, must be computed. Following \cref{sec:formulation:micro}, this is done in two steps: all quantities are first evaluated in the material frame, in which the RVE is actually solved, and subsequently rotated back to the lab frame. For conciseness, the following microscopic quantities, evaluated at the converged solution, are introduced,
\begin{equation}
\bm{F}^* \coloneqq \bm{F}(\bm{w}^*), \qquad \bm{P}^* \coloneqq \bm{P}(\bm{F}^*), \qquad \mathbb{A}^* \coloneqq \mathbb{A}(\bm{F}^*),
\label{eq:starred}
\end{equation}
with $\bm{F}(\bm{w}^*)$ given by \cref{eq:Fmicro} and $\mathbb{A}$ the microscopic tangent stiffness tensor of \cref{eq:lin}.

\paragraph{Macroscopic energy density.}
The macroscopic energy density introduced in \cref{sec:formulation:macro} is the volume average of the microscopic strain energy density $\psi$ evaluated at the converged solution in the material frame,
\begin{equation}
\bar{\Psi}(\bar{\bm{F}}, \bar{v}_i, \bar{\bm{g}}_i) \coloneqq \langle \psi(\bm{F}^*) \rangle.
\label{eq:Psibar}
\end{equation}
Since $\bm{w}^*$ satisfies the equilibrium of \cref{eq:weak} and its sensitivity with respect to any macroscopic input is itself an admissible test function, the implicit contribution of $\bm{w}^*$ to the derivatives of $\bar{\Psi}$ vanishes, and the work-conjugate quantities $\bar{\bm{P}} = \partial\bar{\Psi}/\partial\bar{\bm{F}}$, $\bar{\Gamma}_i = \partial\bar{\Psi}/\partial \bar{v}_i$ and $\bar{\bm{\Lambda}}_i = \partial\bar{\Psi}/\partial\bar{\bm{g}}_i$ of \cref{eq:macro_weak} reduce to the plain volume averages stated below; see~\citet{Rokos2019} for the detailed derivation.

\paragraph{Effective quantities in the material frame.}
Recall from \cref{sec:formulation:micro} that the RVE is solved in the material frame; we denote the quantities obtained from that solve with a tilde. The effective stress, and the mode-conjugate scalar and vector quantities, are then given by
\begin{align}
\tilde{\bm{P}} &\coloneqq \langle \bm{P}^* \rangle, \label{eq:Pbar}\\
\tilde{\Gamma}_i &\coloneqq \langle \bm{P}^* : (\nabla\bm{\phi}_i)^T \rangle, \label{eq:Gamma}\\
\tilde{\bm{\Lambda}}_i &\coloneqq \langle (\bm{P}^*)^T\cdot\bm{\phi}_i + \bm{X}\, (\bm{P}^* : (\nabla\bm{\phi}_i)^T) \rangle. \label{eq:Lambda}
\end{align}
Here $\tilde{\bm{P}}$ is the material-frame counterpart of the effective first Piola--Kirchhoff stress, while $\tilde{\Gamma}_i$ and $\tilde{\bm{\Lambda}}_i$ are work-conjugate to $\bar{v}_i$ and $\bar{\bm{g}}_i$, respectively, consistent with the macroscopic weak form. Note that, because the independent components of the symmetric $\bar{\bm{U}}$ are $\{\bar{U}_{pq}\}_{p\leq q}$, the derivative of $\bar{\Psi}$ with respect to an off-diagonal component ($p<q$) equals $\tilde{P}_{pq} + \tilde{P}_{qp}$ rather than $\tilde{P}_{pq}$; this factor is carried automatically by the basis tensors $\bm{S}^{(pq)}$ introduced below.

\textit{Remark.} In the lab frame, the deformation gradient of the converged solution is $\bar{\bm{R}}\cdot\bm{F}^*$, see \cref{sec:formulation:micro}, and, by objectivity of $\psi$, the corresponding stress is $\bar{\bm{R}}\cdot\bm{P}^*$. The two-point tensor $\bm{P}^*$ thus rotates with $\bar{\bm{R}}$ in its spatial leg. The same holds for the co-rotated modes $\bar{\bm{R}}\cdot\bm{\phi}_i$ and their gradients $\bar{\bm{R}}\cdot(\nabla\bm{\phi}_i)^T$, which replace $\bm{\phi}_i$ and $(\nabla\bm{\phi}_i)^T$ in the lab-frame counterparts of \cref{eq:Gamma,eq:Lambda}. Since the spatial legs are contracted with each other and $\bar{\bm{R}}^T\cdot\bar{\bm{R}} = \bm{I}$, the rotation cancels in the mode-conjugate quantities, but not in the stress,
\begin{align*}
\bar{\Gamma}_i &= \big\langle (\bar{\bm{R}}\cdot\bm{P}^*) : (\bar{\bm{R}}\cdot(\nabla\bm{\phi}_i)^T) \big\rangle = \big\langle \bm{P}^* : (\nabla\bm{\phi}_i)^T \big\rangle = \tilde{\Gamma}_i,\\
\bar{\bm{\Lambda}}_i &= \big\langle (\bar{\bm{R}}\cdot\bm{P}^*)^T\cdot(\bar{\bm{R}}\cdot\bm{\phi}_i) + \bm{X}\, \big((\bar{\bm{R}}\cdot\bm{P}^*) : (\bar{\bm{R}}\cdot(\nabla\bm{\phi}_i)^T)\big) \big\rangle\\
&= \big\langle (\bm{P}^*)^T\cdot\bm{\phi}_i + \bm{X}\, \big(\bm{P}^* : (\nabla\bm{\phi}_i)^T\big) \big\rangle = \tilde{\bm{\Lambda}}_i,\\
\bar{\bm{P}} &= \big\langle \bar{\bm{R}}\cdot\bm{P}^* \big\rangle = \bar{\bm{R}}\cdot\tilde{\bm{P}}.
\end{align*}
Hence $\bar{\Gamma}_i$ and $\bar{\bm{\Lambda}}_i$ are independent of $\bar{\bm{R}}$, whereas the effective stress retains the rotation in its spatial leg, exactly as for the classical (non-enriched) case.

\paragraph{Sensitivity problem.}
Since the microscopic problem is solved at every macroscopic integration point, its tangent must be evaluated efficiently, i.e., without resorting to finite differences. Let $\mu$ denote any scalar macroscopic input, $\mu \in \{\bar{U}_{pq}\}_{p \leq q} \cup \{\bar{v}_i\} \cup \{\bar{g}_{i,k}\}$, where $\{\bar{U}_{pq}\}_{p\leq q}$ denotes the $d(d+1)/2$ independent components of the symmetric stretch and $\bar{g}_{i,k}$ the $k$-th component of $\bar{\bm{g}}_i$. Differentiating \cref{eq:Fmicro}, evaluated in the material frame (i.e., with $\bar{\bm{R}} = \bm{I}$ and $\bar{\bm{F}}$ replaced by $\bar{\bm{U}}$), with respect to $\mu$ at fixed $\bm{w}^*$ gives the explicit sensitivity $\bm{M}_\mu \coloneqq \partial\bm{F}^*/\partial\mu|_{\bm{w}^*}$,
\begin{equation}
\bm{M}_{\bar{U}_{pq}} = \bm{S}^{(pq)}, \qquad \bm{M}_{\bar{v}_i} = (\nabla\bm{\phi}_i)^T, \qquad \bm{M}_{\bar{g}_{i,k}} = \bm{\phi}_i \otimes \bm{e}_k + X_k (\nabla\bm{\phi}_i)^T,
\label{eq:Mmu}
\end{equation}
with $\bm{S}^{(pq)} \coloneqq \bm{e}_p\otimes\bm{e}_q + \bm{e}_q\otimes\bm{e}_p$ for $p<q$ and $\bm{S}^{(pp)} \coloneqq \bm{e}_p\otimes\bm{e}_p$ the symmetric basis tensor associated with the independent stretch component $(p,q)$. The full sensitivity of $\bm{F}^*$ to $\mu$ additionally requires the implicit dependence through $\bm{w}^*$, captured by the sensitivity field $\bm{q}_\mu \coloneqq \partial\bm{w}^*/\partial\mu$, obtained by differentiating the weak form of \cref{eq:weak} at the converged solution,
\begin{equation}
\int_\Omega (\nabla\delta\bm{w})^T : \mathbb{A}^* : (\nabla\bm{q}_\mu)^T\, \mathrm{d}\bm{X} = -\int_\Omega (\nabla\delta\bm{w})^T : \mathbb{A}^* : \bm{M}_\mu\, \mathrm{d}\bm{X} \qquad \forall\, \delta\bm{w},
\label{eq:sens}
\end{equation}
subject to the same periodicity and constraints as \cref{eq:c1,eq:c2,eq:c3}, enforced through Lagrange multipliers $\mathbf{m}_\mu$ in exactly the way $\mathbf{m}$ enforces them in \cref{eq:newton}. Discretizing \cref{eq:sens} thus yields a linear system with the same bordered coefficient matrix as the last Newton iteration of \cref{eq:newton},
\begin{equation}
\begin{bmatrix} \mathbf{K}^* & \mathbf{C}^T \\ \mathbf{C} & \mathbf{0} \end{bmatrix}
\begin{bmatrix} \mathbf{q}_\mu \\ \mathbf{m}_\mu \end{bmatrix}
=
\begin{bmatrix} -\mathbf{b}_\mu \\ \mathbf{0} \end{bmatrix},
\label{eq:sens_discrete}
\end{equation}
where $\mathbf{K}^*$ is assembled from $\mathbb{A}^*$ and the load vector $\mathbf{b}_\mu$ from the right-hand side of \cref{eq:sens}, with $\bm{M}_\mu$ given by \cref{eq:Mmu}. Since this coefficient matrix is already factorized at convergence of \cref{eq:newton}, every additional direction $\mu$ costs only one back-substitution, rather than a full nonlinear solve.

Given $\bm{q}_\mu$, the total sensitivity $\mathrm{d}\bm{F}^*/\mathrm{d}\mu = \bm{M}_\mu + (\nabla\bm{q}_\mu)^T$ yields the derivatives of the material-frame effective quantities through the chain rule,
\begin{align}
\frac{\mathrm{d}\tilde{\bm{P}}}{\mathrm{d}\mu} &= \Big\langle \mathbb{A}^* : \frac{\mathrm{d}\bm{F}^*}{\mathrm{d}\mu} \Big\rangle, \label{eq:dPbar}\\
\frac{\mathrm{d}\tilde{\Gamma}_i}{\mathrm{d}\mu} &= \Big\langle \Big(\mathbb{A}^* : \frac{\mathrm{d}\bm{F}^*}{\mathrm{d}\mu}\Big) : (\nabla\bm{\phi}_i)^T \Big\rangle, \label{eq:dGamma}\\
\frac{\mathrm{d}\tilde{\bm{\Lambda}}_i}{\mathrm{d}\mu} &= \Big\langle \Big(\mathbb{A}^* : \frac{\mathrm{d}\bm{F}^*}{\mathrm{d}\mu}\Big)^T\cdot\bm{\phi}_i + \bm{X}\, \Big(\Big(\mathbb{A}^* : \frac{\mathrm{d}\bm{F}^*}{\mathrm{d}\mu}\Big) : (\nabla\bm{\phi}_i)^T\Big) \Big\rangle. \label{eq:dLambda}
\end{align}

\paragraph{Reconstruction in the lab frame.}
The macroscopic quantities and their tangents with respect to the actual $\bar{\bm{F}}$ are finally recovered from the material-frame quantities via the polar decomposition $\bar{\bm{F}} = \bar{\bm{R}}\cdot\bar{\bm{U}}$; recall from the discussion above \cref{eq:Mmu} that $\mu$ contains components of $\bar{\bm{U}}$, not of $\bar{\bm{F}}$. Since $\bar{\Gamma}_i$ and $\bar{\bm{\Lambda}}_i$ are rotation-invariant, their $\bar{\bm{F}}$-derivatives follow from \cref{eq:dGamma,eq:dLambda} by the chain rule through $\bar{\bm{U}}(\bar{\bm{F}})$ alone,
\begin{equation}
\frac{\mathrm{d}\bar{\Gamma}_i}{\mathrm{d}\bar{F}_{kl}} = \sum_{p \leq q} \frac{\mathrm{d}\tilde{\Gamma}_i}{\mathrm{d}\bar{U}_{pq}}\, \frac{\partial \bar{U}_{pq}}{\partial \bar{F}_{kl}}, \qquad \frac{\mathrm{d}\bar{\bm{\Lambda}}_i}{\mathrm{d}\bar{F}_{kl}} = \sum_{p \leq q} \frac{\mathrm{d}\tilde{\bm{\Lambda}}_i}{\mathrm{d}\bar{U}_{pq}}\, \frac{\partial \bar{U}_{pq}}{\partial \bar{F}_{kl}}.
\label{eq:reconstruct_scalar}
\end{equation}
The effective stress additionally picks up the rotation of $\bar{\bm{R}}$ itself,
\begin{equation}
\frac{\mathrm{d}\bar{\bm{P}}}{\mathrm{d}\bar{F}_{kl}} = \frac{\partial \bar{\bm{R}}}{\partial \bar{F}_{kl}}\cdot \tilde{\bm{P}} + \sum_{p \leq q} \bar{\bm{R}}\cdot \frac{\mathrm{d}\tilde{\bm{P}}}{\mathrm{d}\bar{U}_{pq}}\, \frac{\partial \bar{U}_{pq}}{\partial \bar{F}_{kl}}.
\label{eq:reconstruct_Pbar}
\end{equation}
In \cref{eq:reconstruct_scalar,eq:reconstruct_Pbar} the sums run over the $d(d+1)/2$ independent components of $\bar{\bm{U}}$ only, consistent with the definition of $\bm{S}^{(pq)}$ in \cref{eq:Mmu}; the summation convention does not apply to $p$ and $q$. The closed-form expressions for $\partial\bar{\bm{R}}/\partial\bar{\bm{F}}$ and $\partial\bar{\bm{U}}/\partial\bar{\bm{F}}$, obtained from the eigendecomposition of $\bar{\bm{C}} \coloneqq \bar{\bm{F}}^T\cdot\bar{\bm{F}}$, are standard and omitted here for brevity; see, e.g.,~\citep{Chen1993}. Derivatives with respect to $\bar{v}_i$ and $\bar{\bm{g}}_i$ do not involve $\bar{\bm{R}}$, since these inputs leave the polar decomposition of $\bar{\bm{F}}$ unaffected; they are recovered from \cref{eq:dPbar,eq:dGamma,eq:dLambda} directly, with $\mathrm{d}\bar{\bm{P}}/\mathrm{d}\bar{v}_i = \bar{\bm{R}}\cdot \mathrm{d}\tilde{\bm{P}}/\mathrm{d}\bar{v}_i$ (and similarly for $\bar{\bm{g}}_i$), while $\bar{\Gamma}_i$ and $\bar{\bm{\Lambda}}_i$ pass through unchanged.

\section{Reduced order model}\label{sec:rom}

The microscopic problem of \cref{sec:formulation:micro}, together with the sensitivity problems supplying its tangent, has to be solved at every macroscopic integration point of \cref{eq:macro_weak}, in every Newton iteration and in every load step. This section develops the ROM for it, in two stages. The fluctuation field is first projected onto a low-dimensional basis obtained by POD, which reduces the number of unknowns (\cref{sec:rom:pod}). The remaining dependence on the full finite element mesh, through the integrals that have to be evaluated at every iteration, is then removed by hyperreduction, i.e., by replacing the full quadrature with a small set of integration points and weights selected once in an offline stage (\cref{sec:rom:hyperreduction}).

\subsection{Proper orthogonal decomposition}\label{sec:rom:pod}

To reduce the number of DOFs, the fluctuation field $\bm{w}$ is approximated with a reduced basis,
\begin{equation}
\bm{w} \approx \sum_{n=1}^{N_r} a_n \bm{w}_n,
\label{eq:podansatz}
\end{equation}
where $N_r$ is the number of retained modes, typically $N_r \ll N_{\mathrm{dof}}$ with $N_{\mathrm{dof}}$ the number of DOFs of the discretized fluctuation field. The fluctuation modes $\{\bm{w}_n\}_{n=1}^{N_r}$ are obtained by applying POD to a set of pre-computed snapshots of $\bm{w}^*$ for different combinations of the macroscopic inputs $(\bar{\bm{U}}, \bar{v}_i, \bar{\bm{g}}_i)$; the POD is carried out with respect to the $H^1$ inner product. Since every snapshot satisfies the periodicity and constraints of \cref{eq:c1,eq:c2,eq:c3}, and since these are linear and homogeneous in $\bm{w}$, any linear combination of snapshots --- and hence any $\bm{w}$ represented by \cref{eq:podansatz} --- satisfies them automatically.

\textit{Remark.} Unlike the full-order problem, the constraints of \cref{eq:c1,eq:c2,eq:c3} therefore do not need to be enforced explicitly in the reduced order problem, which removes the Lagrange-multiplier block from \cref{eq:newton} altogether and allows for more efficient, unconstrained linear solvers.

Inserting \cref{eq:podansatz} into \cref{eq:weak,eq:lin} and assuming a Galerkin projection, the reduced internal force $\mathbf{f}^{\mathrm{r}} \in \mathbb{R}^{N_r}$ and reduced tangent stiffness $\mathbf{K}^{\mathrm{r}} \in \mathbb{R}^{N_r\times N_r}$ follow as
\begin{equation}
f^{\mathrm{r}}_n(\mathbf{a}) \coloneqq \int_\Omega (\nabla\bm{w}_n)^T : \bm{P}(\bm{F}(\mathbf{a}))\, \mathrm{d}\bm{X}, \qquad K^{\mathrm{r}}_{nm}(\mathbf{a}) \coloneqq \int_\Omega (\nabla\bm{w}_n)^T : \mathbb{A}(\bm{F}(\mathbf{a})) : (\nabla\bm{w}_m)^T\, \mathrm{d}\bm{X},
\label{eq:fK_reduced}
\end{equation}
for $n,m = 1,\dots,N_r$, where $\mathbf{a} = [a_1,\dots,a_{N_r}]^T$ collects the unknown reduced coefficients, solved for with Newton's method analogous to \cref{eq:newton} but without the Lagrange-multiplier block. Again in the material frame, $\bm{F}(\mathbf{a})$ follows from \cref{eq:Fmicro} upon inserting \cref{eq:podansatz},
\begin{equation}
\bm{F}(\mathbf{a}) = \bar{\bm{U}} + \sum_{i=1}^{N_\phi}\Big[\bm{\phi}_i\otimes\bar{\bm{g}}_i + (\bar{v}_i + \bm{X}\cdot\bar{\bm{g}}_i)(\nabla\bm{\phi}_i)^T\Big] + \Big(\sum_{n=1}^{N_r} a_n \nabla\bm{w}_n\Big)^T.
\label{eq:Fmicro_reduced}
\end{equation}

\textit{Remark.} The effective quantities and their tangents are computed exactly as for the full-order model in \cref{sec:formulation:effective}, only in the reduced space: in the reduced counterpart of the sensitivity problem of \cref{eq:sens_discrete}, the (already factorized) reduced stiffness $\mathbf{K}^{\mathrm{r}}$ of \cref{eq:fK_reduced} replaces the bordered coefficient matrix (the constraints being satisfied by construction), the right-hand sides built from $\bm{M}_\mu$ are projected onto the reduced basis, and each direction $\mu$ again costs a single back-substitution, now of a small dense $N_r \times N_r$ system. The derivatives of \cref{eq:dPbar,eq:dGamma,eq:dLambda} are then evaluated with $\mathrm{d}\bm{F}^*/\mathrm{d}\mu = \bm{M}_\mu + (\nabla\bm{q}_\mu)^T$, where the sensitivity field $\bm{q}_\mu$ is expanded in the reduced basis. The reduced order computation thus parallels the full-order one step by step.

\subsection{Hyperreduction}\label{sec:rom:hyperreduction}

While the reduced system of \cref{eq:fK_reduced} involves only $N_r$ unknowns, evaluating the integrals therein, as well as the effective quantities $\tilde{\bm{P}}$, $\tilde\Gamma_i$ and $\tilde{\bm{\Lambda}}_i$ of \cref{eq:Pbar,eq:Gamma,eq:Lambda} and their sensitivity-based tangents of \cref{sec:formulation:effective}, still requires integration over the entire microstructural finite element mesh. To accelerate this, we seek a small subset of the original integration points, together with corresponding non-negative weights, that closely reproduces the full quadrature of every quantity that must remain accurate: the reduced internal force $f^{\mathrm{r}}_n(\mathbf{a})$, the effective stress $\tilde{\bm{P}}$, and the mode-conjugate quantities $\tilde\Gamma_i$ and $\tilde{\bm{\Lambda}}_i$.

The stress field itself is first compressed with POD~\citep{Hernandez2014},
\begin{equation}
\bm{P} \approx \sum_{m=1}^{M_r} b_m \bm{\chi}_m,
\label{eq:stress_pod}
\end{equation}
where $\{\bm{\chi}_m\}_{m=1}^{M_r}$ are $L^2$-orthonormal stress modes obtained from stress snapshots and $M_r$ is the number of retained stress modes. Inserting \cref{eq:stress_pod} into \cref{eq:fK_reduced,eq:Pbar,eq:Gamma,eq:Lambda} writes all four target quantities out explicitly,
\begin{align}
f^{\mathrm{r}}_n &\approx \sum_{m=1}^{M_r} b_m \int_\Omega (\nabla\bm{w}_n)^T : \bm{\chi}_m\, \mathrm{d}\bm{X}, &
\tilde{\bm{P}} &\approx \sum_{m=1}^{M_r} b_m \big\langle \bm{\chi}_m \big\rangle, \label{eq:targets_stress}\\
\tilde\Gamma_i &\approx \sum_{m=1}^{M_r} b_m \big\langle \bm{\chi}_m : (\nabla\bm{\phi}_i)^T \big\rangle, &
\tilde{\bm{\Lambda}}_i &\approx \sum_{m=1}^{M_r} b_m \big\langle \bm{\chi}_m^T\cdot\bm{\phi}_i + \bm{X}\,\big(\bm{\chi}_m : (\nabla\bm{\phi}_i)^T\big) \big\rangle. \label{eq:targets_modes}
\end{align}
Every integral on the right-hand sides is a fixed number, assembled once offline from the stress modes and the precomputed fields $\nabla\bm{w}_n$, $\bm{\phi}_i$, $(\nabla\bm{\phi}_i)^T$ and $\bm{X}$; the coefficients $b_m$ are the only part that varies, changing with the macroscopic input $(\bar{\bm{U}}, \bar{v}_i, \bar{\bm{g}}_i)$ as the RVE is driven along a load path. Provided \cref{eq:stress_pod} approximates \emph{any} stress field the RVE problem produces, the reduced integration rule must therefore integrate each mode-weighted integrand accurately in its own right, i.e., for an arbitrary choice of the $b_m$; all four target quantities then follow at once.

Evaluating these integrands at all $N_Q$ points of the full quadrature and collecting them row-wise yields the target matrix and its fully integrated right-hand side,
\begin{equation}
\mathbf{A} = \begin{bmatrix} \mathbf{A}_1 \\ \mathbf{A}_2 \\ \mathbf{A}_3 \\ \mathbf{A}_4 \\ \mathbf{A}_5 \end{bmatrix} \in \mathbb{R}^{N_{\mathrm{t}} \times N_Q}, \qquad
\mathbf{b} \coloneqq \mathbf{A}\,\bm{\omega} \in \mathbb{R}^{N_{\mathrm{t}}},
\label{eq:target_matrix}
\end{equation}
where $A_{rq}$ is the $r$-th target integrand evaluated at the $q$-th integration point of the full quadrature and $\bm{\omega} \in \mathbb{R}^{N_Q}$ collects the full quadrature weights. The right-hand side $\mathbf{b}$ thus holds exactly the fully integrated value of every target quantity. The five blocks hold, respectively, the integrands of the reduced internal force $f^{\mathrm{r}}_n$ ($\mathbf{A}_1$), of the effective stress $\tilde{\bm{P}}$ ($\mathbf{A}_2$), of $\tilde{\bm{\Lambda}}_i$ ($\mathbf{A}_3$) and of $\tilde\Gamma_i$ ($\mathbf{A}_4$), and a row of ones ($\mathbf{A}_5$), whose entry in $\mathbf{b}$ is the RVE volume $|\Omega|$ and which therefore forces the reduced rule to preserve it.
The total number of rows follows directly from \cref{eq:targets_stress,eq:targets_modes} as
\begin{equation}
N_{\mathrm{t}} = \underbrace{N_r M_r}_{\mathbf{A}_1} + \underbrace{d^2 M_r}_{\mathbf{A}_2} + \underbrace{d\,N_\phi M_r}_{\mathbf{A}_3} + \underbrace{N_\phi M_r}_{\mathbf{A}_4} + \underbrace{1}_{\mathbf{A}_5}.
\label{eq:Nt}
\end{equation}

Each of the $N_Q$ columns of $\mathbf{A}$ corresponds to one integration point of the full quadrature. Constructing the reduced rule therefore amounts to retaining only $N_q \ll N_Q$ of those columns, collected in $\mathbf{A}^{\mathrm{r}} \in \mathbb{R}^{N_{\mathrm{t}} \times N_q}$, and assigning them new non-negative weights $\bm{\omega}^{\mathrm{r}} \in \mathbb{R}^{N_q}$, such that the retained columns alone still reproduce $\mathbf{b}$,
\begin{equation}
\mathbf{A}^{\mathrm{r}}\,\bm{\omega}^{\mathrm{r}} \approx \mathbf{b}, \qquad \bm{\omega}^{\mathrm{r}} \geq \bm{0}, \qquad N_q \ll N_Q.
\label{eq:ecm_problem}
\end{equation}


The integration points and non-negative weights are selected with the same greedy, empirical-cubature-inspired algorithm as in our previous work~\citep{Guo2025}, itself building on the ECM of \citet{Hernandez2017}, which incrementally selects columns of $\mathbf{A}$ so as to minimize a non-negative least-squares residual; we refer to~\citet{Guo2025} for the full details and do not repeat it here. The greedy selection is terminated either once the relative residual, which characterizes the mismatch between the reduced and the full quadrature of all target quantities, drops below a user-prescribed tolerance $\varepsilon_{\mathrm{ECM}}$, or if a selected target number $N_q$ of the reduced integration rule has been reached.

\subsection{Summary of the offline and online stages}\label{sec:rom:summary}

For clarity, the complete procedure is collected here.

\paragraph{Offline stage.}
The offline stage is executed once per microstructure and consists of four steps:
\begin{enumerate}
\item extract the enrichment modes $\bm{\phi}_i$, obtained here by driving the RVE towards its first bifurcation and taking the critical eigenmode(s) (\cref{sec:examples:2d,sec:examples:3d});
\item sample the macroscopic inputs $(\bar{\bm{U}}, \bar{v}_i, \bar{\bm{g}}_i)$ within physically motivated bounds and solve the full-order microscopic problem of \cref{eq:newton} along a load ramp for each sample, storing the converged fluctuation and stress fields;
\item compute the displacement basis $\{\bm{w}_n\}_{n=1}^{N_r}$ by $H^1$-POD of the fluctuation snapshots, and the stress basis $\{\bm{\chi}_m\}_{m=1}^{M_r}$ by $L^2$-POD;
\item assemble the target matrix of \cref{eq:target_matrix} and run the greedy selection to obtain the $N_q$ integration points and their non-negative weights.
\end{enumerate}

\paragraph{Online stage.}
The online stage then replaces the full-order RVE at every macroscopic integration point: given $(\bar{\bm{F}}, \bar{v}_i, \bar{\bm{g}}_i)$, the polar decomposition $\bar{\bm{F}} = \bar{\bm{R}}\cdot\bar{\bm{U}}$ is formed, the reduced problem of \cref{eq:fK_reduced} is solved for $\mathbf{a}\in\mathbb{R}^{N_r}$ by Newton's method with all integrals evaluated over the $N_q$ selected points, one back-substitution per macroscopic input yields the sensitivity fields, and the effective quantities and tangent blocks are assembled from \cref{eq:Pbar,eq:Gamma,eq:Lambda,eq:dPbar,eq:dGamma,eq:dLambda} and rotated back to the lab frame through \cref{eq:reconstruct_scalar,eq:reconstruct_Pbar}.

\section{Numerical examples}\label{sec:examples}

Two examples are considered: a two-dimensional specimen for which both reference solutions remain affordable, so that the implementation and the reduction can each be validated directly (\cref{sec:examples:2d}), and a fully three-dimensional lattice for which neither reference can be computed and the reduced model is what makes the two-scale analysis possible at all (\cref{sec:examples:3d}). Four model classes carry the comparison throughout: the DNS resolving the entire specimen, the micromorphic FE\textsuperscript{2} model with full-order RVEs (MM-FOM), its reduced order counterpart obtained by POD alone (MM-ROM), and the additionally hyperreduced model (MM-HROM).

All four are implemented in the open-source finite element framework DOLFINx (version 0.10.0)~\citep{Baratta2023}. Periodicity is enforced with the dolfinx\_mpc library (version 0.10.5)~\citep{Dokken2025} and the constraints of \cref{eq:c1,eq:c2,eq:c3} by the null-space projection of \cref{sec:formulation:micro}; the eigenvalue problems arising in the stability monitoring are solved with SLEPc (version 3.25.1)~\citep{Hernandez2005}, and linear systems are solved with a sparse direct LU factorization using MUMPS (version 5.8.2)~\citep{Amestoy2001} through PETSc (version 3.25.2)~\citep{Balay2025}. All meshes are generated with Gmsh (version 4.15.2)~\citep{Geuzaine2009}. The library implementing all of the above is available in our \texttt{fe2\_rom} repository.\footnote{\url{https://github.com/theronguo/fe2_rom}}

In all computations reported below, the microscopic Newton iteration of \cref{eq:newton} is terminated at a relative residual tolerance of $10^{-10}$ or an absolute tolerance $\varepsilon_{\mathrm{newton}} = 10^{-8}$, with at most 30 iterations. The macroscopic solve uses a relative tolerance of $10^{-6}$ throughout, and an absolute tolerance of $10^{-6}$. Load stepping is adaptive throughout, halving the increment upon divergence and enlarging it by 1.5x if Newton converges within 5 iterations. Bifurcations are traversed with a perturbation method, employed for both the DNS and the macroscopic micromorphic problems: the eigenvalues of the tangent stiffness matrix are monitored at the converged state of every load step and, upon loss of stability, the solution is perturbed along the critical eigenmode with initially small and successively larger amplitudes until Newton's method converges to a stable state with all eigenvalues positive. All two-dimensional computations of \cref{sec:examples:2d} are carried out with 8 MPI processes on a single node equipped with an AMD EPYC 9474F 48-core processor, while the three-dimensional ones in~\cref{sec:examples:3d} use 27 MPI processes.

\subsection{Two-dimensional example}\label{sec:examples:2d}

\begin{figure}[pos={!ht}]
\centering
\includegraphics[width=\linewidth]{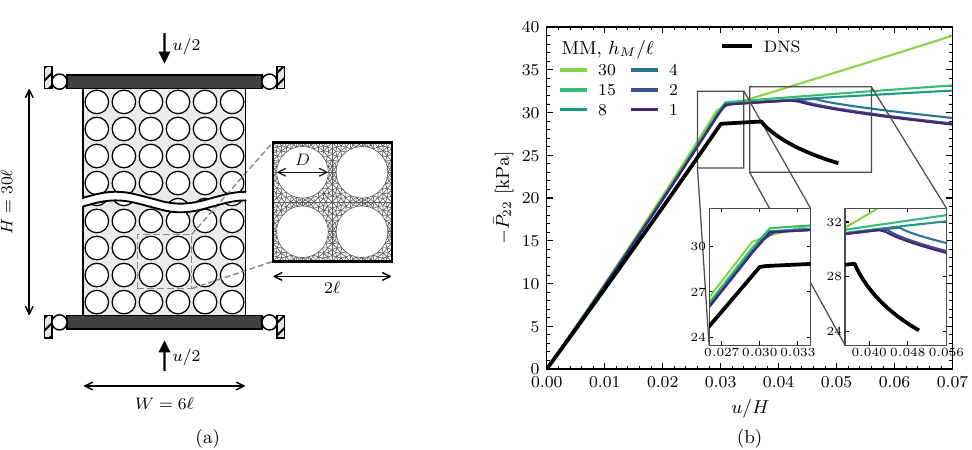}
\caption{Two-dimensional example, after Example~1 of \citet{vanBree2020}. (a) Geometry: the $W=6\ell \times H=30\ell$ specimen (left), compressed through the rigid loading platens. The periodic $2\ell\times2\ell$ RVE mesh (quadratic triangles, $h_{\mathrm{m}} = \ell/10$) is shown on the right. (b) Nominal stress $-\bar{P}_{22}$ vs.\ nominal strain $u/H$: DNS reference (black) and micromorphic FE\textsuperscript{2} solutions for macroscopic mesh sizes $h_{\mathrm{M}}/\ell \in \{1,2,4,8,15,30\}$ (in color); the insets zoom into the first (local patterning) and second (global buckling) bifurcations.}
\label{fig:geometry_p22}
\end{figure}

The first example reproduces Example 1 of \citet{vanBree2020}, chosen because it comes with published DNS and MM-FOM reference data against which our implementation can be validated directly, and because the problem exhibits two bifurcations: local patterning and global buckling. The specimen, drawn in \cref{fig:geometry_p22}(a) (left), is a finite elastomeric column of width $W$ and height $H$ under plane-strain conditions, patterned with a square stacking of unit cells of edge length $\ell = 9.97~\mathrm{mm}$, each containing a circular hole of diameter $D = 8.67~\mathrm{mm}$, following the design of \citet{Bertoldi2008}. Guided by the mesh-convergence study and DNS comparison of~\citet[Fig.\ 5]{vanBree2020}, we consider a specimen of $W = 6\ell$ and $H = 30\ell$ (slenderness ratio $H/W = 5$), compressed up to an overall vertical strain of $u/H = 0.07$, applied through equal and opposite displacements $\pm u/2\,\bm{e}_2$ on the bottom and top edges; these are additionally clamped laterally ($u_1 = 0$) so that they remain straight and do not contract. In the micromorphic computations, the mode amplitude is suppressed along the two loaded edges, i.e., $\bar{v}_1 = 0$ in \cref{eq:macro_bc}, since the stiff loading platens prevent the pattern from developing there, whereas the lateral edges carry the natural boundary conditions of \cref{eq:macro_bc_nat}. The overall response is reported in terms of the macroscopic nominal stress $\bar{P}_{22} \coloneqq \mathcal{R}/W$, where $\mathcal{R}$ denotes the total vertical reaction force along the top edge per unit out-of-plane thickness; since compression corresponds to negative values of $\bar{P}_{22}$, the compressive nominal stress $-\bar{P}_{22}$ is reported throughout.

\Cref{fig:geometry_p22}(a) (right) shows the periodic $2\ell\times2\ell$ RVE used for the microscopic problem of \cref{sec:formulation:micro}, containing four full holes of diameter $D$, one centered in each of the four $\ell\times\ell$ unit cells that tile the RVE; an RVE spanning $2\times2$ unit cells is required because the patterning mode alternates between neighboring holes and is hence $2\ell$-periodic~\citep{Bertoldi2008}. The RVE is discretized with quadratic (P2) triangles of characteristic size $h_{\mathrm{m}} = \ell/10$ (592 elements), matching the DNS discretization, which resolves the entire specimen with 26\,640 quadratic triangles (58\,861 nodes) and six quadrature points per element.

The elastomeric base material is modeled with the compressible hyperelastic law of \citet{Bertoldi2008}, also adopted by \citet{vanBree2020}, with strain energy density
\begin{equation}
\psi(\bm{F}) = c_1 (I_1 - 3) + c_2 (I_1-3)^2 - 2 c_1 \ln J + \tfrac{1}{2} K (J-1)^2,
\label{eq:bertoldi}
\end{equation}
where $J \coloneqq \det\bm{F}$ and $I_1 \coloneqq \operatorname{tr}\bm{C}$ is the first invariant of the right Cauchy--Green tensor $\bm{C} \coloneqq \bm{F}^T\cdot\bm{F}$. The constitutive parameters, taken from the experimental characterization of \citet{Bertoldi2008}, are listed in \cref{tab:bertoldi}.

\begin{table}[pos={!ht}]
\centering
\caption{Constitutive parameters of the hyperelastic law of \cref{eq:bertoldi}, from \citet{Bertoldi2008}.}
\label{tab:bertoldi}
\begin{tabular*}{\tblwidth}{@{}LLL@{}}
\toprule
$c_1$ [MPa] & $c_2$ [MPa] & $K$ [MPa] \\
\midrule
0.55 & 0.3 & 55 \\
\bottomrule
\end{tabular*}
\end{table}

A single patterning mode $\bm{\phi}_1$ emerges for this microstructure ($N_\phi=1$), which we extract numerically as follows. The RVE is driven towards uniaxial strain compression along the loading direction with an incremental Newton procedure, and at every converged state the smallest eigenvalues of the tangent stiffness matrix are computed; as soon as a negative eigenvalue appears, the load step is rejected and the increment halved, which is repeated until a minimal increment is reached. At the resulting converged state, located just below the critical load, the eigenvalue problem is solved once more and the eigenmode associated with the smallest eigenvalue is taken as $\bm{\phi}_1$. For this particular microstructure, a closed-form approximation of the patterning mode is moreover available~\citep[Eq.~(56)]{vanBree2020},
\begin{equation}
\bm{\phi}_1^{\mathrm{cf}}(\bm{X}) \propto -\big(s_+ + s_-\big)\, \bm{e}_1 + \big(s_+ - s_-\big)\, \bm{e}_2, \qquad s_\pm \coloneqq \sin\!\big(\pi (X_2 \pm X_1)/\ell\big),
\label{eq:phi_cf}
\end{equation}
with the origin of $\bm{X}$ at the RVE center. The numerically extracted mode closely matches this expression: quantifying the difference through the normalization- and sign-invariant relative $L^2$ misalignment of two mode fields $\bm{a}$ and $\bm{b}$,
\begin{equation}
E_\phi(\bm{a}, \bm{b}) \coloneqq \min_{\alpha \in \mathbb{R}} \sqrt{\frac{\big\langle \|\bm{a} - \alpha\, \bm{b}\|^2 \big\rangle}{\big\langle \|\bm{a}\|^2 \big\rangle}},
\label{eq:phi_misalignment}
\end{equation}
we find $E_\phi(\bm{\phi}_1, \bm{\phi}_1^{\mathrm{cf}}) = 8.0\times10^{-2}$; the small remaining difference reflects geometric detail of the actual perforated microstructure that the smooth sinusoidal approximation of \cref{eq:phi_cf} does not resolve.

\subsubsection{Validation of implementation}\label{sec:examples:2d:validation}
DNS reference solutions are obtained by fully resolving the specimen with the element type and density stated above. \Cref{fig:geometry_p22}(b) compares the resulting nominal stress $-\bar{P}_{22}$ against $u/H$ with the micromorphic FE\textsuperscript{2} solution for a range of macroscopic mesh sizes $h_{\mathrm{M}}/\ell \in \{1,2,4,8,15,30\}$, analogous to~\citet[Fig.\ 5]{vanBree2020}: the thick black curve is the DNS reference, and the colored curves are the micromorphic FE\textsuperscript{2} solutions, colored from light (coarse, $h_{\mathrm{M}}=30\ell$) to dark (fine, $h_{\mathrm{M}}=1\ell$). The response is initially linear until a first bifurcation near $u/H \approx 0.03$, beyond which local patterning emerges and the stiffness drops sharply. All discretizations capture this first bifurcation, albeit slightly shifted from the DNS reference due to homogenization inaccuracy; see~\citet{Sperling2024} for a detailed numerical study of scale separation in various computational homogenization schemes. Only the finer discretizations $h_{\mathrm{M}}/\ell = 4$, $2$ and $1$, however, capture the second bifurcation, where the specimen buckles globally and the response softens; they place its peak at $u/H = 0.046$, $0.042$ and $0.042$, respectively, against $u/H = 0.037$ for the DNS, so that the residual shift with respect to the reference is a homogenization effect that mesh refinement does not remove. The difference between $h_{\mathrm{M}}/\ell = 2$ and $1$ is marginal --- $0.09$~kPa in RMS over the whole loading, i.e., $0.3\%$ of the peak stress --- so the macroscopic mesh can be regarded as converged. The DNS itself is terminated at $u/H = 0.05$, where opposing hole boundaries of the buckled microstructure come into contact, which the model does not account for; the micromorphic simulations do not resolve the individual hole boundaries and are therefore continued up to $u/H = 0.07$, so that both bifurcations are fully traversed. This is consistent with the mesh-convergence behavior reported in~\citep{vanBree2020} and confirms that our implementation reproduces their published results.

\subsubsection{Reduced order model}\label{sec:examples:2d:rom}
Having verified the micromorphic implementation against the DNS, we next assess the proposed ROM. All reduced order results reported below are obtained for the macroscopic discretization $h_{\mathrm{M}} = 4\ell$, the coarsest discretization that still captures both bifurcations; the macroscopic mesh then consists of 32 quadratic triangles with three quadrature points each, i.e., 96 RVE problems are solved per macroscopic residual or tangent evaluation. The corresponding MM-FOM solution serves as the reference throughout. Three offline choices are studied, in this order: the number of retained POD modes $N_r$, the number of training load trajectories (64 and 128), and the number $N_q$ of retained integration points.

The training snapshots are generated by sampling the macroscopic inputs $(\bar{\bm{U}}, \bar{v}_1, \bar{\bm{g}}_1)$ within physically motivated bounds; only the symmetric stretch needs to be sampled, i.e., three independent components in two dimensions. The normal components are sampled within $\bar{U}_{11}, \bar{U}_{22} \in [0.90, 1.05]$ and the shear component within $\bar{U}_{12} \in [-0.05, 0.05]$. The largest compressive deviation from identity, $\varepsilon_{\max} = 0.10$, covers the applied overall strain of $u/H = 0.07$ with some margin, while the moderate tensile and shear ranges account for the local deformation states visited by individual macroscopic quadrature points. The amplitude bound is chosen such that the average pattern displacement over the RVE, $|\bar{v}_1|\, \langle \|\bm{\phi}_1\| \rangle$, may reach a prescribed fraction $\kappa$ of the maximal average cell shortening $\varepsilon_{\max}\, \ell_{\mathrm{RVE}}$, i.e., $|\bar{v}_1| \leq \kappa\,\varepsilon_{\max}\, \ell_{\mathrm{RVE}} / \langle \|\bm{\phi}_1\| \rangle$ with $\ell_{\mathrm{RVE}} = 2\ell$ the RVE edge length, a definition that is invariant with respect to the normalization of $\bm{\phi}_1$; the gradient bound follows per component as $|\bar{g}_{1,k}| \leq |\bar{v}_1|_{\max} / \ell_{\mathrm{RVE}}$, where $|\bar{v}_1|_{\max}$ denotes the amplitude bound just defined. In the present example we take $\kappa = 1$. 

From the resulting six-dimensional box, 64 end points are drawn with Latin hypercube sampling, and each end point is reached by a linear ramp from the reference configuration, $(\bar{\bm{U}}(t), \bar{v}_1(t), \bar{\bm{g}}_1(t)) = (\bm{I} + t\, (\bar{\bm{U}} - \bm{I}),\, t\, \bar{v}_1,\, t\, \bar{\bm{g}}_1)$ with $t \in [0, 1]$, solved with the adaptive load stepping described above, attempting the full ramp in a single increment first and allowing the increment to shrink to $10^{-5}$. Every accepted increment yields a converged microscopic state, so that a single load trajectory generally contributes multiple snapshots. For this example, all 64 trajectories converged and produced 1297 snapshots of the fluctuation field $\bm{w}^*$ and the stress field $\bm{P}^*$.

\begin{figure}[pos={!ht}]
\centering
\includegraphics[width=\linewidth]{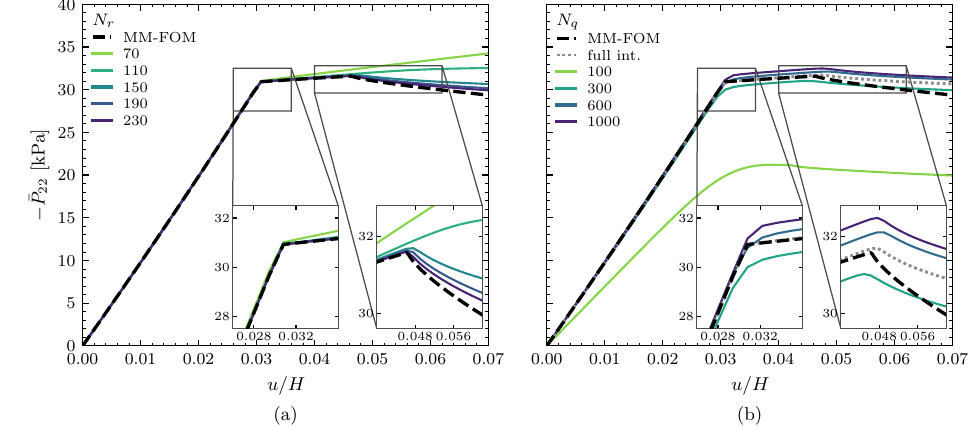}
\caption{Reduced order results at $h_{\mathrm{M}} = 4\ell$: nominal stress $-\bar{P}_{22}$ vs.\ nominal strain $u/H$ of the MM-ROM compared against the MM-FOM reference. (a) POD study under full integration for basis sizes $N_r \in \{70, 110, 150, 190, 230\}$ (64-trajectory training set). (b) ECM study at fixed $N_r = M_r = 150$ for reduced integration rules with $N_q \in \{100, 300, 600, 1000\}$ points, with the full-integration curve of (a) included as the $N_q \rightarrow 3552$ reference. The insets zoom into the first (local patterning) and second (global buckling) bifurcations.}
\label{fig:p22_rom_studies}
\end{figure}

To isolate the error of the POD--Galerkin projection from that of the hyperreduction, the MM-ROM is studied first, i.e., \cref{eq:fK_reduced} and the effective quantities are integrated with the full quadrature rule of the underlying mesh, so that the only approximation is the reduced basis of dimension $N_r$. The displacement basis is obtained by POD of the 1297 fluctuation snapshots with respect to the $H^1$ inner product. As a measure of the approximation quality of the basis itself, we consider the relative best-approximation error of the training set,
\begin{equation}
E(N_r) \coloneqq \sqrt{\frac{\sum_{n > N_r} \lambda_n}{\sum_{n} \lambda_n}},
\label{eq:best_approx}
\end{equation}
where $\lambda_1 \geq \lambda_2 \geq \dots$ are the POD eigenvalues, i.e., the relative $H^1$ error of projecting the snapshots onto the leading $N_r$ modes; the quantity $1 - E^2$ is the commonly reported fraction of snapshot energy captured. \Cref{tab:best_approx} lists $E(N_r)$ for the basis sizes $N_r \in \{70, 110, 150, 190, 230\}$ considered: the error decays smoothly and near-exponentially by more than two orders of magnitude over that range, without any visible threshold. \Cref{fig:p22_rom_studies}(a) compares the resulting $-\bar{P}_{22}$ curves against the MM-FOM. To quantify the deviation from the reference, the root-mean-square (RMS) error
\begin{equation}
E_{\mathrm{RMS}} \coloneqq \Bigg(\frac{1}{n_s}\sum_{k=1}^{n_s} \Big[P^{\mathrm{ROM}}(u_k/H) - P^{\mathrm{ref}}(u_k/H)\Big]^2\Bigg)^{1/2}
\label{eq:rms}
\end{equation}
is employed, where $P$ denotes the reported nominal stress ($-\bar{P}_{22}$ here and $-\bar{P}_{zz}$ in \cref{sec:examples:3d}) and $u_k/H$, $k = 1,\dots,n_s$, are the load levels of the reference solution, onto which the reduced curve is interpolated.

For $N_r \leq 110$, the MM-ROM overshoots the buckling load and misses the second bifurcation entirely: the response keeps rising monotonically to the end of the loading, with $E_{\mathrm{RMS}} = 1.91$ and $1.25$~kPa for $N_r = 70$ and $110$, respectively. From $N_r = 150$ onwards, both bifurcations are captured: the peak load matches the MM-FOM to within $0.4\%$ and its location to within a single load increment, and $E_{\mathrm{RMS}}$ drops to $0.50$~kPa, i.e., $1.6\%$ of the peak stress, decreasing further to $0.30$~kPa at $N_r = 190$ and $0.20$~kPa at $N_r = 230$. The residual error is concentrated in the deep post-buckling branch: up to the peak, every basis with $N_r \geq 150$ reproduces the reference to within $0.1$~kPa, whereas at $u/H = 0.07$ the $N_r = 150$ model lies $1.30$~kPa ($4.4\%$) above the MM-FOM, which reduces to $0.56$~kPa ($1.9\%$) at $N_r = 230$.

\begin{table}[pos={!ht}]
\centering
\caption{POD study at $h_{\mathrm{M}} = 4\ell$ under full integration: relative best-approximation error $E(N_r)$ of \cref{eq:best_approx} ($H^1$ norm, displacement fluctuation snapshots) and error $E_{\mathrm{RMS}}$ of \cref{eq:rms} of the macroscopic $-\bar{P}_{22}$ curve with respect to the MM-FOM, for increasing basis size $N_r$ and for the two training sets of 64 and 128 load trajectories. The models marked with a dagger do not reach the second bifurcation.}
\label{tab:best_approx}
\begin{tabular*}{\tblwidth}{@{}L@{\extracolsep{\fill}}LLLLL@{}}
\toprule
$N_r$ & 70 & 110 & 150 & 190 & 230 \\
\midrule
$E(N_r)$, 64 trajectories & $5.4\times10^{-3}$ & $1.3\times10^{-3}$ & $3.4\times10^{-4}$ & $1.0\times10^{-4}$ & $3.4\times10^{-5}$ \\
$E_{\mathrm{RMS}}$ [kPa], 64 trajectories & 1.91$^\dagger$ & 1.25$^\dagger$ & 0.50 & 0.30 & 0.20 \\
\midrule
$E(N_r)$, 128 trajectories & $1.1\times10^{-2}$ & $3.8\times10^{-3}$ & $1.6\times10^{-3}$ & $7.3\times10^{-4}$ & $3.6\times10^{-4}$ \\
$E_{\mathrm{RMS}}$ [kPa], 128 trajectories & 1.86$^\dagger$ & 1.35$^\dagger$ & 0.63 & 0.26 & 0.15 \\
\bottomrule
\end{tabular*}
\end{table}

To study the representativeness of the training data, the training set was subsequently enriched with 64 additional Latin hypercube samples, resulting in 128 trajectories and 2578 snapshots. The results remain, however, largely similar, see the lower half of \cref{tab:best_approx}. The smallest basis capturing both bifurcations is still $N_r = 150$, and the macroscopic error is of the same size throughout, marginally larger at $N_r = 150$ ($0.63$ against $0.50$~kPa) and marginally smaller at $N_r = 230$ ($0.15$ against $0.20$~kPa). It is noted that $E(N_r)$ is uniformly larger for the enriched set, simply because a richer snapshot pool is harder to represent with the same number of modes. This reflects a general difficulty of sampling-based training for problems with bifurcations: the loading trajectories actually traversed by the macroscopic problem are not known a priori without solving the two-scale problem itself, which would defeat the purpose of constructing a ROM in the first place. Since doubling the training set buys neither a smaller basis nor a systematically more accurate response, the 64-trajectory set is retained in what follows, so that the POD and the hyperreduction study below are built from the same snapshots.

\textit{Remark.} We also experimented with a Hermite-type basis, in which the snapshot set is enriched with the sensitivity fields $\partial\bm{w}^*/\partial\mu$ that are available from the sensitivity problem of \cref{eq:sens_discrete} at negligible extra cost. This consistently led to considerably larger bases without a significant improvement of the macroscopic response; a sufficiently rich set of plain fluctuation snapshots evidently already spans the required sensitivity directions.

Next, the hyperreduction error is studied. Based on the previous study, the displacement and stress bases are fixed at $N_r = 150$ (64 trajectories) and $M_r = 150$ modes ($L^2$ inner product). We prescribe the number $N_q$ of retained points directly rather than a tolerance $\varepsilon_{\mathrm{ECM}}$, since $N_q$ is the quantity that governs the online cost, and terminate the greedy selection of \cref{sec:rom:hyperreduction} at $N_q \in \{100, 300, 600, 1000\}$. The candidate set from which the points are selected is the six-point (degree-four) quadrature rule also employed by the full-order solver, 3552 points in total. \Cref{tab:ecm_points} reports, for each rule, the relative ECM residual attained at termination and the error $E_{\mathrm{RMS}}$ of the resulting $-\bar{P}_{22}$ curve with respect to the MM-FOM, and \cref{fig:p22_rom_studies}(b) shows the corresponding curves.

The coarsest rule, $N_q = 100$ ($2.8\%$ of the candidate points), is entirely inadequate: the response is far too compliant already before the first bifurcation, underestimating the MM-FOM by $35\%$ at $u/H = 0.03$, and it reaches a peak of only $21.2$~kPa against $31.6$~kPa for the reference. From $N_q = 300$ ($8.4\%$) onwards, both bifurcations are recovered and the hyperreduced curves follow the reference closely: the pre-buckling branch is reproduced to within $2.2\%$ at $N_q = 300$ and to within $0.7\%$ at $N_q = 600$ and $1000$, and the peak load is captured to within $1.8\%$, $1.7\%$ and $2.9\%$, respectively. The corresponding $E_{\mathrm{RMS}}$ values, $0.37$, $0.82$ and $1.02$~kPa, are of the same order as the projection error of the underlying basis itself ($0.50$~kPa at the same $N_r$, cf.\ \cref{tab:best_approx}). Measured instead against its own full-integration counterpart (MM-ROM), which isolates the hyperreduction error, the three rules deviate by $0.55$, $0.35$ and $0.57$~kPa. That the error with respect to the MM-FOM happens to be smallest for $N_q = 300$ is a coincidence: its slightly too compliant integration rule partly cancels the stiffening introduced by the POD projection, whereas for $N_q = 600$ and $1000$ the two contributions add.

Since all three rules reproduce the reference to within a few percent and the accuracy does not improve with further points, $N_q = 300$, i.e., $8.4\%$ of the original quadrature points, is adopted as the best trade-off between integration cost and induced error for this example.

\begin{table}[pos={!ht}]
\centering
\caption{ECM study at $N_r = M_r = 150$: prescribed number $N_q$ of selected integration points (out of 3552 candidate quadrature points), the relative ECM residual attained at termination, and the error $E_{\mathrm{RMS}}$ of \cref{eq:rms} with respect to the MM-FOM and to the fully integrated $N_r = 150$ model; full integration is included for reference. The $N_q = 100$ rule marked with a dagger does not reproduce either bifurcation.}
\label{tab:ecm_points}
\begin{tabular*}{\tblwidth}{@{}L@{\extracolsep{\fill}}LLLLL@{}}
\toprule
$N_q$ & 100 & 300 & 600 & 1000 & 3552 \\
\midrule
fraction of full [\%] & 2.8 & 8.4 & 16.9 & 28.2 & 100 \\
ECM residual & $5.4\times10^{-2}$ & $1.5\times10^{-2}$ & $5.9\times10^{-3}$ & $2.5\times10^{-3}$ & --- \\
$E_{\mathrm{RMS}}$ vs.\ MM-FOM [kPa] & 8.57$^\dagger$ & 0.37 & 0.82 & 1.02 & 0.50 \\
$E_{\mathrm{RMS}}$ vs.\ MM-ROM [kPa] & 8.78$^\dagger$ & 0.55 & 0.35 & 0.57 & --- \\
\bottomrule
\end{tabular*}
\end{table}

\subsubsection{Computational cost}\label{sec:examples:2d:cost}

Finally, we comment on the computational cost. \Cref{tab:timings_2d} collects the measured wall-clock times of the four model classes at $h_{\mathrm{M}} = 4\ell$, all run with 8 MPI processes on the hardware stated at the beginning of \cref{sec:examples}, with the MM-ROM taken at $N_r = 150$ and the MM-HROM at $N_r = M_r = 150$ and $N_q = 300$. Since the DNS terminates at $u/H = 0.05$ whereas the micromorphic models are continued to $0.07$, all times are those needed to reach $u/H = 0.05$, the largest strain the four have in common, so that they measure the same load path.

The reduction with respect to the MM-FOM is substantial: in the online stage the MM-ROM is roughly $46$ times and the MM-HROM roughly $85$ times faster, so that a two-scale computation that takes an hour in its full-order form completes in well under a minute. All speed-ups quoted below are online speed-ups in this sense, the offline stage being reported separately at the end of this section. The origin of the gain is structural. Per Newton iteration of a single RVE solve, the full-order model evaluates the constitutive law at 3552 integration points (six per element) and solves the sparse constrained system of \cref{eq:newton} with roughly 2750 displacement unknowns, plus one sensitivity back-substitution per macroscopic input for the tangents. The MM-ROM solves instead a dense system of size $N_r = 150$, which alone accounts for most of the observed factor of $46$, while the MM-HROM in addition loops over only 300 rather than 3552 integration points, which removes the remaining mesh-dependent work and yields the further factor of $1.9$. This second factor is much smaller than the ratio of point counts, since a part of the online cost, in particular the assembly and the repeated solution of the small dense systems, is independent of $N_q$.

The comparison with the DNS is far less dramatic, as expected. In two dimensions the DNS is itself comparatively inexpensive --- it is a single nonlinear solve --- whereas every micromorphic FE\textsuperscript{2} step, reduced or not, must solve a microscopic problem at each of the 96 macroscopic quadrature points. Over the same load path the MM-HROM is about $3.5$ times faster than the DNS, i.e., the two remain of the same order. Whenever the DNS is affordable, which is typically the case in two dimensions, it therefore remains the method of choice; the benefit of the MM-HROM materializes when the DNS becomes computationally intractable, as illustrated by the three-dimensional example of \cref{sec:examples:3d}.

It should be noted that these timings must be read with some caution, as neither our DNS nor our two-scale implementation is optimized, and the measured values depend on implementation and solver details; the DNS cost, in particular, is dominated by the perturbation strategy used to traverse the bifurcations and by the settings of the employed eigenvalue solvers.

Regarding the offline stage, each RVE simulation is run serially on a single core, with the samples distributed over the available 8 cores. The extraction of $\bm{\phi}_1$ together with the 64 training solves takes $2.8$~min, applying the POD and the greedy point selection took $12$~s and $34$~s. In total, the offline stage takes around $3.6$~min.

\begin{table}[pos={!ht}]
\centering
\caption{Two-dimensional example at $h_{\mathrm{M}} = 4\ell$: measured wall-clock times to reach $u/H = 0.05$, the largest strain traversed by all four models, on 8 MPI processes of a single AMD EPYC 9474F 48-core node. The MM-ROM uses $N_r = 150$ and the MM-HROM $N_r = M_r = 150$ with $N_q = 300$. Online speed-ups are with respect to the MM-FOM; they exclude the offline stage, whose cost is quantified in the text.}
\label{tab:timings_2d}
\begin{tabular*}{\tblwidth}{@{}L@{\extracolsep{\fill}}LLLL@{}}
\toprule
 & DNS & MM-FOM & MM-ROM & MM-HROM \\
\midrule
microscopic unknowns per RVE & --- & 2746 & 150 & 150 \\
microscopic integration points per RVE & --- & 3552 & 3552 & 300 \\
wall time to $u/H = 0.05$ [s] & 146 & 3574 & 78 & 42 \\
online speed-up vs.\ MM-FOM & 24 & 1 & 46 & 85 \\
\bottomrule
\end{tabular*}
\end{table}

\subsection{Three-dimensional example}\label{sec:examples:3d}

\begin{figure}[pos={!ht}]
\centering
\includegraphics[width=\linewidth]{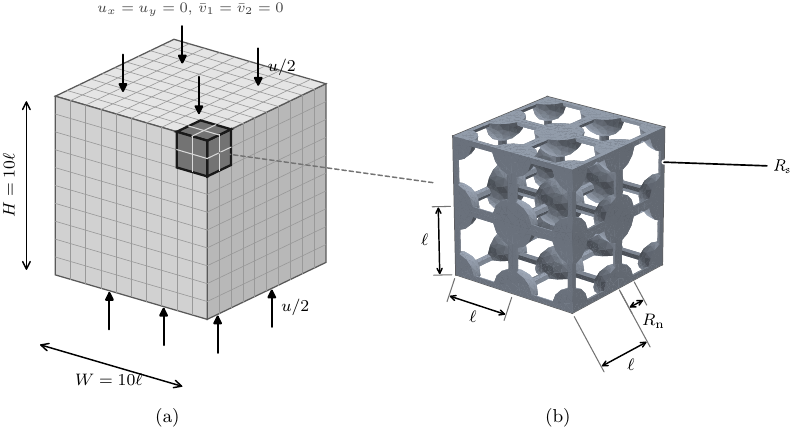}
\caption{Three-dimensional example, after the simple-cubic lattice metamaterial of \citet{Meijer2026}. (a) The macroscopic specimen: a cube of $10\times10\times10$ unit cells, compressed along $z$ by equal and opposite displacements $\pm u/2$ on the two loaded faces, which are additionally held flat ($u_x = u_y = 0$) and on which the mode amplitudes are suppressed ($\bar{v}_1 = \bar{v}_2 = 0$); the four lateral faces are traction-free. (b) The periodic $2\times2\times2$-cell sphere-strut RVE, called out from the specimen: spherical nodes of radius $R_{\mathrm{n}}$ on the cubic-cell corners (edge $\ell$) are joined by slender struts of radius $R_{\mathrm{s}}$, and uniaxial compression along $z$ buckles the load-aligned struts.}
\label{fig:geometry_rve_3d}
\end{figure}

The second example is fully three-dimensional. We consider the simple-cubic lattice metamaterial of \citet{Meijer2026}: a spherical node of radius $R_{\mathrm{n}}$ is centered on every corner of a cubic unit cell of edge length $\ell$, and neighboring nodes are joined along the cell edges by slender circular struts of radius $R_{\mathrm{s}}$, see \cref{fig:geometry_rve_3d}. In contrast to \citet{Meijer2026}, who assume the spheres to be rigid, the nodes are here made of the same material as the struts. Throughout this example the Cartesian axes $1,2,3$ are labeled $x,y,z$ and are aligned with the cell edges, with $z$ the loading direction. Under uniaxial compression the load-aligned struts carry the load and buckle; as demonstrated experimentally and numerically by \citet{Meijer2026}, the strut slenderness $R_{\mathrm{s}}/\ell$ selects the buckling character, ranging from a global, Euler-type buckling of the whole specimen for very slender struts to a local, pattern-transforming instability in which neighboring struts buckle in alternating directions. To promote the local patterning regime, we take $R_{\mathrm{n}} = \ell/3$ and $R_{\mathrm{s}} = \ell/14$ (very thin struts), and model the base material with the compressible hyperelastic law of \cref{eq:bertoldi} and the parameters of \cref{tab:bertoldi}. As in the two-dimensional example, the microscopic RVE comprises $2\times2\times2$ unit cells (\cref{fig:geometry_rve_3d}(b)), which accommodates the period-doubled patterning mode; periodic boundary conditions are imposed on its fluctuation field.

The macroscopic specimen, shown in \cref{fig:geometry_rve_3d}(a), is a cube of $10\times10\times10$ unit cells, with $W = H = 10\ell$. It is compressed along $z$ by equal and opposite displacements $\pm u/2\,\bm{e}_z$ applied on the two loaded faces $z = 0$ and $z = H$, up to an overall strain of $u/H = 0.05$, i.e., to roughly three times the strain at which the struts buckle. As in \cref{sec:examples:2d}, the loaded faces are clamped laterally ($u_x = u_y = 0$) so that they remain flat and do not contract, whereas the four lateral faces are traction-free; in the micromorphic computations the mode amplitudes are likewise suppressed on the loaded faces, $\bar{v}_1 = \bar{v}_2 = 0$ in \cref{eq:macro_bc}, with the natural boundary conditions of \cref{eq:macro_bc_nat} on the remainder. The overall response is again reported as the compressive nominal stress $-\bar{P}_{zz}$, with $\bar{P}_{zz} \coloneqq \mathcal{R}/W^2$ and $\mathcal{R}$ the total vertical reaction force on the top face; the homogenized RVE stress of \cref{sec:examples:3d:convergence} is reported in the same way.

Whereas in two dimensions the DNS of the full specimen remained affordable (\cref{sec:examples:2d}), in three dimensions it becomes intractable: resolving the curved sphere-strut geometry costs orders of magnitude more degrees of freedom per unit cell, and the repeated tangent factorizations demanded by the stability eigenproblem and by the eigenmode-perturbation strategy used to traverse the bifurcations make a direct simulation prohibitively expensive already for a handful of cells in each direction. The micromorphic FE\textsuperscript{2} model (MM-FOM) removes the need to resolve the full specimen, but with a full-order RVE solved at every macroscopic integration point it remains very expensive; the reduced order model is what renders the two-scale computation feasible within a reasonable time, and is the object of this example. Both statements are made quantitative in \cref{sec:examples:3d:convergence}, once the microscopic discretization has been fixed.

\subsubsection{Microscopic mesh convergence}\label{sec:examples:3d:convergence}

Before turning to the two-scale computations, we establish the microscopic discretization needed to resolve the RVE response accurately, since every subsequent result inherits this resolution. To isolate the discretization error, we consider the RVE alone under uniaxial strain compression, $\bar{\bm{F}} = \operatorname{diag}(1,1,\lambda)$ with $\lambda$ decreasing from $1$ to $0.90$, i.e., up to $10\%$ compression, solved with the full-order classical (first-order) homogenization scheme, i.e., the formulation of \cref{sec:formulation} without enrichment modes ($N_\phi=0$), and periodic boundary conditions; the buckling (patterning) occurs at $1-\bar F_{zz} \approx 0.017$. The curved geometry is discretized with quadratic (P2) tetrahedra, whose iso-parametric curved facets are required to represent the spherical nodes and, above all, the cylindrical struts whose bending governs the buckling response, and integrated with a $4$ integration points per tetrahedron.

\begin{figure}[pos={!ht}]
\centering
\includegraphics[width=0.5\linewidth]{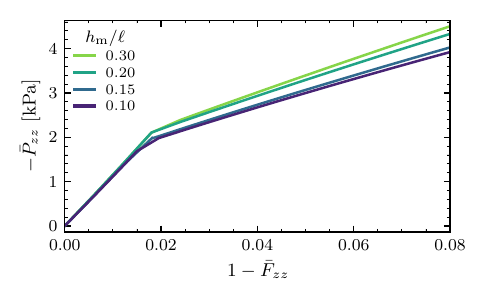}
\caption{Microscopic mesh convergence for the three-dimensional RVE under uniaxial strain compression (classical first-order homogenization, periodic boundary conditions, degree-$2$ quadrature), homogenized nominal stress $-\bar P_{zz}$ vs.\ compressive strain $1-\bar F_{zz}$, shown up to $8\%$ (below the second buckling), for element sizes $h_{\mathrm{m}}/\ell \in \{0.30, 0.20, 0.15, 0.10\}$.}
\label{fig:pzz_mesh_qp}
\end{figure}

We vary the characteristic element size $h_{\mathrm{m}}/\ell \in \{0.30,\, 0.20,\, 0.15,\, 0.10\}$. \Cref{fig:pzz_mesh_qp} shows the resulting homogenized nominal stress $-\bar P_{zz}$ versus the compressive strain $1-\bar F_{zz}$. All discretizations coincide in the pre-buckling regime and reproduce the buckling onset; the post-buckling branch, by contrast, softens with every refinement, since finer meshes represent the strut bending more compliantly. Relative to the finest mesh ($h_{\mathrm{m}}/\ell = 0.10$, about $9.1 \times 10^4$ displacement DOF), the root-mean-square deviation of the stress curve over the loading range drops from $7.1\%$ at $h_{\mathrm{m}}/\ell = 0.30$ to $5.5\%$ at $0.20$ and $1.3\%$ at $0.15$. We therefore adopt $h_{\mathrm{m}}/\ell = 0.15$ in the following. To quantify the remaining discretization error more precisely, we split the deviation at the buckling onset: the adopted mesh lies within $0.3\%$ of the finest one before buckling but $1.4\%$ after it, so that the post-buckling branch is, by a factor of roughly four, the slowest part of the response to converge; at the end of the loading range its stress is still $2.9\%$ high.

This split also delimits what the adopted mesh may be claimed to resolve. The pre-buckling stiffness and the buckling onset are converged, and it is these that govern the two-scale response reported below. The post-buckling branch, by contrast, softens monotonically with every refinement and is not fully converged even at the finest mesh we could afford, the peak stress still dropping by about $5\%$ from $h_{\mathrm{m}}/\ell = 0.15$ to $0.10$; resolving it would require a finer mesh at a cost we judged disproportionate for a discretization that must subsequently be solved at every macroscopic integration point. The post-buckling stresses reported in \cref{sec:examples:3d:rom} therefore carry a systematic bias of this order, which exceeds the reduction errors quoted there and should be kept in mind when reading them.

With the resulting configuration (a $2\times2\times2$-cell RVE discretized with quadratic tetrahedra at $h_{\mathrm{m}}/\ell = 0.15$ and degree-$2$ quadrature) the microscopic problem has $57\,819$ displacement degrees of freedom and $36\,864$ quadrature points ($9216$ elements with $4$ points each). This fixes the size of the full-order RVE problem that must be solved, or reduced, at every macroscopic integration point in the two-scale computations that follow.

At this resolution, neither of the two reference solutions used in \cref{sec:examples:2d} could be computed with the resources at our disposal. A DNS of the $10\times10\times10$-cell specimen at the same $h_{\mathrm{m}}/\ell = 0.15$ comprises $2.29$ million nodes, i.e., roughly $6.9$ million displacement degrees of freedom, each load step of which requires a Newton solve, followed by the stability eigenproblem of the perturbation strategy at the converged state. The micromorphic FE\textsuperscript{2} model with full-order RVEs is out of reach for a different reason: at the macroscopic discretization adopted below it carries $648$ macroscopic quadrature points, each of which requires, per macroscopic Newton iteration and per load step, a constrained Newton solve of the $57\,819$-unknown RVE problem of \cref{eq:newton} followed by fourteen sensitivity back-substitutions of \cref{eq:sens_discrete} --- one for each of the six independent components of $\bar{\bm{U}}$, for $\bar{v}_1$ and $\bar{v}_2$ of the two patterning modes introduced in \cref{sec:examples:3d:rom}, and for the six components of $\bar{\bm{g}}_1$ and $\bar{\bm{g}}_2$. The accuracy of the reduced model can therefore not be assessed against a full-order reference in this example. Instead, we assess it by internal convergence: the number $N_r$ of retained POD modes and, subsequently, the number $N_q$ of retained integration points are increased systematically until the macroscopic response ceases to change, which is the subject of \cref{sec:examples:3d:rom}.

\subsubsection{Reduced order model}\label{sec:examples:3d:rom}

We now turn to the two-scale micromorphic computations. Three ingredients have to be fixed before the two-scale results can be presented: the enrichment modes $\bm{\phi}_i$, the training set from which the reduced model is built, and the macroscopic discretization. All are addressed in turn below, after which the reduced basis and the reduced integration rule are studied in the same two stages as in \cref{sec:examples:2d}.

\begin{figure}[pos={!ht}]
\centering
\includegraphics[width=0.93\linewidth]{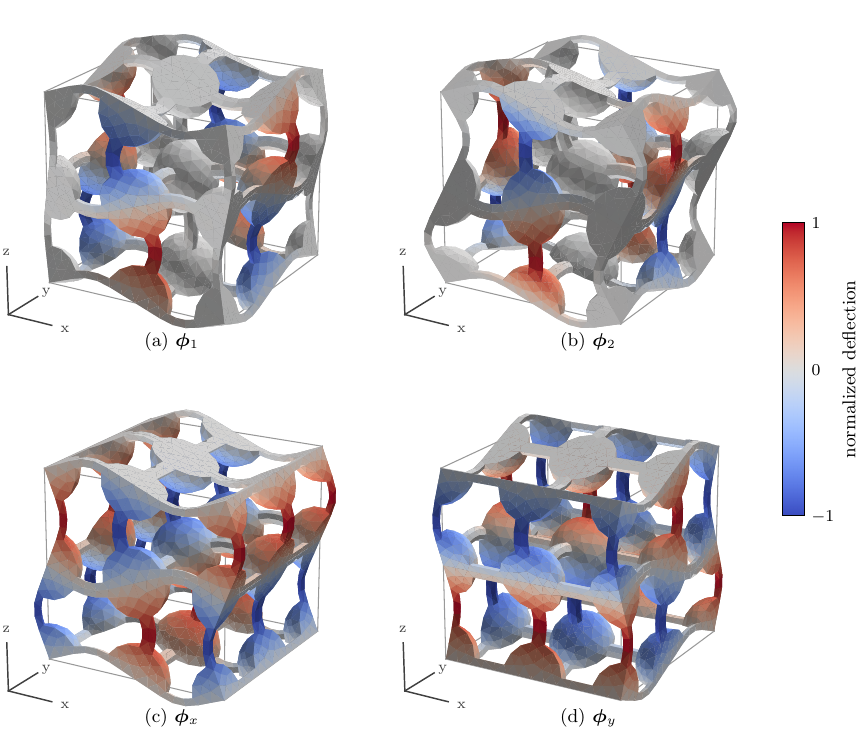}
\caption{Three-dimensional example: the degenerate patterning eigenspace obtained at the near-critical uniaxial-compression state, visualized as the RVE surface warped by each mode and colored by the mode's displacement component along the in-plane direction in which the load-aligned struts deflect, i.e., along $(\bm{e}_x+\bm{e}_y)/\sqrt{2}$, $(\bm{e}_x-\bm{e}_y)/\sqrt{2}$, $\bm{e}_x$ and $\bm{e}_y$ in (a)--(d), respectively, normalized by its maximum magnitude in each panel. (a) The eigenvector $\bm{\phi}_1$ returned by the solver and (b) its orthogonal partner $\bm{\phi}_2$; in these the load-aligned struts deflect along the two in-plane diagonals. (c) $\bm{\phi}_x = (\bm{\phi}_1+\bm{\phi}_2)/\sqrt{2}$ and (d) $\bm{\phi}_y = (\bm{\phi}_1-\bm{\phi}_2)/\sqrt{2}$, the polarized combinations spanning the same eigenspace, which are retained as the enrichment modes, and in which the struts deflect purely along $x$ and along $y$, respectively. In all four the deflection reverses sign between neighboring cells both in-plane and along $z$, which is the period doubling the $2\times2\times2$-cell RVE accommodates.}
\label{fig:phi_modes_3d}
\end{figure}

\begin{figure}[pos={!ht}]
\centering
\includegraphics[width=0.5\linewidth]{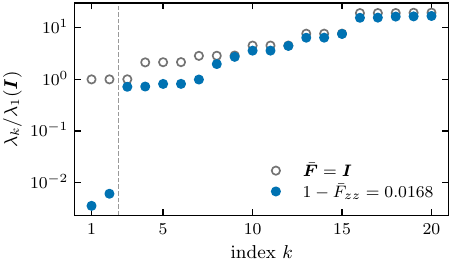}
\caption{Three-dimensional example: the $20$ smallest eigenvalues $\lambda_k$ of the tangent stiffness matrix of the RVE, normalized by the smallest one of the undeformed RVE, at the undeformed state (open circles) and at the near-critical uniaxial-compression state at which the enrichment modes are extracted (filled circles). The dashed line marks the gap below which the degenerate pair $\bm{\phi}_1$, $\bm{\phi}_2$ of \cref{fig:phi_modes_3d}(a) and (b) lies.}
\label{fig:phi_spectrum_3d}
\end{figure}

The enrichment modes are extracted with the same procedure as in \cref{sec:examples:2d}, here with the RVE driven towards uniaxial strain compression along the loading axis, $\bar{\bm{F}} = \operatorname{diag}(1,1,1-\bar F_{zz})$. The procedure arrests at $1 - \bar F_{zz} = 0.0168$, just below the first bifurcation (cf.\ \cref{sec:examples:3d:convergence}), where the eigenvalue problem is solved once more. In contrast to the two-dimensional example, two eigenmodes $\bm{\phi}_1$ and $\bm{\phi}_2$ are found below the first gap in the spectrum (\cref{fig:phi_modes_3d}(a) and (b)), and they are degenerate to within the resolution of the computation. As \cref{fig:phi_spectrum_3d} shows, their eigenvalues drop by more than two orders of magnitude towards the critical state, whereas all others decrease by at most a factor of three, so that a clear gap separates the pair from the remainder of the spectrum. The degeneracy follows directly from the symmetry of the loading: compression along $z$ leaves the microstructure invariant under an exchange of $x$ and $y$, so that the two in-plane deflection directions of the load-aligned struts carry the same critical load. Indeed, the second mode is the image of the first under a rotation by $90^\circ$ about the loading axis: the misalignment of \cref{eq:phi_misalignment} between $\bm{\phi}_2$ and the rotated $\bm{\phi}_1$ evaluates to $3.2\times10^{-2}$.

The two-dimensional eigenspace spanned by the pair is described most transparently by the orthogonal combinations $\bm{\phi}_x \coloneqq (\bm{\phi}_1 + \bm{\phi}_2)/\sqrt{2}$ and $\bm{\phi}_y \coloneqq (\bm{\phi}_1 - \bm{\phi}_2)/\sqrt{2}$, shown in \cref{fig:phi_modes_3d}(c) and (d): in each of them the load-aligned struts deflect purely along one in-plane direction, $x$ or $y$ respectively, reversing their deflection between neighboring cells both along that direction and along $z$.

We retain the whole eigenspace, hence $N_\phi = 2$, so that the in-plane direction of the pattern is left to the macroscopic solution. Of the four fields discussed above we take the polarized combinations $\bm{\phi}_x$ and $\bm{\phi}_y$ (\cref{fig:phi_modes_3d}(c) and (d)) rather than the raw eigenvectors $\bm{\phi}_1$ and $\bm{\phi}_2$, since they are the physically interpretable single-plane patterns, see~\cite{Meijer2026}. In the following we hence write $\bm{\phi}_1 \coloneqq \bm{\phi}_x$ and $\bm{\phi}_2 \coloneqq \bm{\phi}_y$ for the retained modes.

The training snapshots are generated by sampling the macroscopic inputs $(\bar{\bm{U}}, \bar{v}_i, \bar{\bm{g}}_i)$. We choose to sample the normal components within $\bar U_{xx}, \bar U_{yy}, \bar U_{zz} \in [0.90, 1.02]$ and the shear components within $\bar U_{xy}, \bar U_{xz}, \bar U_{yz} \in [-0.02, 0.02]$. The amplitude bound follows the same definition as in \cref{sec:examples:2d}, $|\bar{v}_i| \leq \kappa\, \varepsilon_{\max}\, \ell_{\mathrm{RVE}} / \langle \|\bm{\phi}_i\| \rangle$ with $\ell_{\mathrm{RVE}} = 2\ell$, and $|\bar{g}_{i,k}| \leq |\bar{v}_i|_{\max}/\ell_{\mathrm{RVE}}$ per component, and we choose $\kappa = 0.5$.

From the resulting fourteen-dimensional box --- six stretch components, the amplitudes $\bar{v}_1$ and $\bar{v}_2$, and the six components of $\bar{\bm{g}}_1$ and $\bar{\bm{g}}_2$ --- 256 end points are drawn with Latin hypercube sampling and reached by a linear ramp from the reference configuration, solved with an adaptive load stepping, every accepted increment of which contributes one snapshot of $\bm{w}^*$ and $\bm{P}^*$. Of the 256 trajectories, 232 reach their end point; the remainder stall at a microscopic instability and contribute the increments traversed before stalling. The pool comprises 3664 snapshots in total.

All two-scale results below are obtained on a macroscopic mesh of element size $h_{\mathrm{M}} \approx 3.3\ell$, i.e., $3\times3\times3$ boxes subdivided into 162 quadratic tetrahedra with four quadrature points each, so that 648 microscopic problems are solved per macroscopic residual or tangent evaluation. The macroscopic system itself is small; the cost of the two-scale computation resides entirely in these 648 RVEs.

\begin{figure}[pos={!ht}]
\centering
\includegraphics[width=\linewidth]{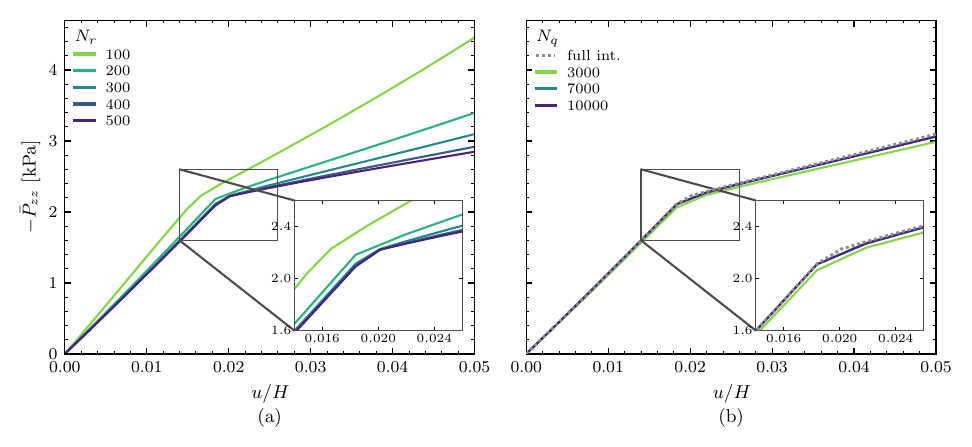}
\caption{Reduced order results for the three-dimensional example at $h_{\mathrm{M}} \approx 3.3\ell$: nominal stress $-\bar{P}_{zz}$ vs.\ nominal strain $u/H$. (a) POD study under full integration (MM-ROM) for basis sizes $N_r \in \{100, 200, 300, 400, 500\}$. (b) ECM study at fixed $N_r = M_r = 300$ for reduced integration rules with $N_q \in \{3000, 7000, 10\,000\}$ selected points, with the full-integration curve of (a) included as the $N_q \rightarrow 36\,864$ reference. The insets zoom into the buckling knee.}
\label{fig:pzz_rom_studies}
\end{figure}

As in \cref{sec:examples:2d}, the error of the POD--Galerkin projection is first isolated from that of the hyperreduction by integrating \cref{eq:fK_reduced} and the effective quantities with the full quadrature rule of the underlying mesh, so that the reduced basis of dimension $N_r$ is the only approximation. The displacement basis is obtained by POD of the fluctuation snapshots with respect to the $H^1$ inner product, and the basis sizes $N_r \in \{100, 200, 300, 400, 500\}$ are considered. \Cref{fig:pzz_rom_studies}(a) shows the corresponding $-\bar{P}_{zz}$ curves and \cref{tab:best_approx_3d} the associated errors. All models reproduce the same qualitative response: an initially linear branch up to $u/H \approx 0.018$, at which point the load-aligned struts buckle and the tangent stiffness drops by a factor of about five, followed by a shallower, rising post-buckling branch. A linear response up to the onset of buckling followed by a marked loss of stiffness is also measured by \citet[Fig.~2C]{Meijer2026} on their lattices, although for their most slender struts the force levels off to a plateau rather than continuing to rise.

Quantitatively, and taking the most resolved model with $N_r = 500$ as the reference, $N_r = 100$ is clearly insufficient: it overestimates the pre-buckling stiffness by $22.2\%$ and the terminal stress by $56.5\%$. From $N_r = 200$ onwards the pre-buckling branch and the buckling onset are captured --- the initial secant stiffness agrees with the reference to within $4.3\%$ at $N_r = 200$, $1.0\%$ at $N_r = 300$ and $0.2\%$ at $N_r = 400$ --- and mainly the post-buckling branch still drifts downwards with increasing basis size, the terminal stress ending $19.2\%$ above the reference at $N_r = 200$, $8.7\%$ above it at $N_r = 300$ and $2.5\%$ above it at $N_r = 400$. Correspondingly, the error $E_{\mathrm{RMS}}$ of \cref{eq:rms} falls monotonically from $0.243$~kPa at $N_r = 200$ to $0.105$~kPa at $N_r = 300$ and $0.029$~kPa at $N_r = 400$. As a compromise between accuracy and cost, we build the hyperreduced models on $N_r = 300$, which captures the pre-buckling branch and the buckling onset to within $1\%$ and the post-buckling branch to within $9\%$; the hyperreduction study below is not affected by this choice, since it measures the deviation from the fully integrated model with the same basis.

\begin{table}[pos={!ht}]
\centering
\caption{Three-dimensional POD study under full integration: relative best-approximation error $E(N_r)$ of \cref{eq:best_approx} ($H^1$ norm, displacement fluctuation snapshots) and error $E_{\mathrm{RMS}}$ of \cref{eq:rms} of the macroscopic $-\bar{P}_{zz}$ curve for increasing basis size $N_r$, together with the deviations of its pre-buckling secant stiffness and its terminal stress. The deviations are taken with respect to the most resolved model, $N_r = 500$.}
\label{tab:best_approx_3d}
\begin{tabular*}{\tblwidth}{@{}L@{\extracolsep{\fill}}LLLLL@{}}
\toprule
$N_r$ & 100 & 200 & 300 & 400 & 500 \\
\midrule
$E(N_r)$ & $6.4\times10^{-2}$ & $2.2\times10^{-2}$ & $9.2\times10^{-3}$ & $4.2\times10^{-3}$ & $2.1\times10^{-3}$ \\
pre-buckling stiffness dev.\ [\%] & $+22.2$ & $+4.3$ & $+1.0$ & $+0.2$ & --- \\
terminal stress dev.\ [\%] & $+56.5$ & $+19.2$ & $+8.7$ & $+2.5$ & --- \\
$E_{\mathrm{RMS}}$ [kPa] & 0.748 & 0.243 & 0.105 & 0.029 & --- \\
\bottomrule
\end{tabular*}
\end{table}

The hyperreduction error is studied next. The displacement basis is fixed at $N_r = 300$, and the stress basis is likewise truncated at $M_r = 300$ modes ($L^2$ inner product). In total, there are $36\,864$ candidate points on $9216$ elements. As in \cref{sec:examples:2d:rom}, we directly choose the number $N_q$ of retained points rather than prescribing $\varepsilon_{\mathrm{ECM}}$, and construct hyperreduced models with $N_q \in \{3000, 7000, 10\,000\}$, i.e., between $8.1\%$ and $27.1\%$ of the candidate points.

\begin{table}[pos={!ht}]
\centering
\caption{Three-dimensional ECM study at $N_r = M_r = 300$: prescribed number $N_q$ of selected integration points (out of $36\,864$ candidate quadrature points), the relative ECM residual attained at termination, and the error $E_{\mathrm{RMS}}$ of \cref{eq:rms} with respect to the full-integration $N_r = 300$ model; full integration is included for reference.}
\label{tab:ecm_points_3d}
\begin{tabular*}{\tblwidth}{@{}L@{\extracolsep{\fill}}LLLL@{}}
\toprule
$N_q$ & 3000 & 7000 & 10\,000 & 36\,864 \\
\midrule
fraction of full [\%] & 8.1 & 19.0 & 27.1 & 100 \\
ECM residual & $7.7\times10^{-2}$ & $2.8\times10^{-2}$ & $1.6\times10^{-2}$ & --- \\
$E_{\mathrm{RMS}}$ [kPa] & 0.062 & 0.017 & 0.020 & --- \\
\bottomrule
\end{tabular*}
\end{table}

\Cref{fig:pzz_rom_studies}(b) shows the resulting $-\bar{P}_{zz}$ curves and \cref{tab:ecm_points_3d} reports $E_{\mathrm{RMS}}$ with respect to the full-integration $N_r = 300$ reference. All three rules traverse the loading. The coarsest rule, $N_q = 3000$ ($8.1\%$ of the candidate points), already reproduces the entire response qualitatively --- the pre-buckling branch, the buckling onset and the post-buckling slope --- but is visibly too compliant, underestimating the pre-buckling stiffness by $2.0\%$ and the terminal stress by $3.5\%$, and gives $E_{\mathrm{RMS}} = 0.062$~kPa, i.e., $2.0\%$ of the terminal stress. At $N_q = 7000$ ($19.0\%$) the deviation drops by a factor of almost four to $0.017$~kPa ($0.6\%$), with the pre-buckling stiffness matched to $0.1\%$, and remains at this level at $N_q = 10\,000$ ($27.1\%$, $0.020$~kPa). Roughly one fifth of the original quadrature points therefore suffices to render the hyperreduction error small against the projection error of the underlying basis, and we adopt $N_q = 7000$ as the best trade-off between integration cost and induced error.

\begin{figure}[pos={!ht}]
\centering
\includegraphics[width=\linewidth]{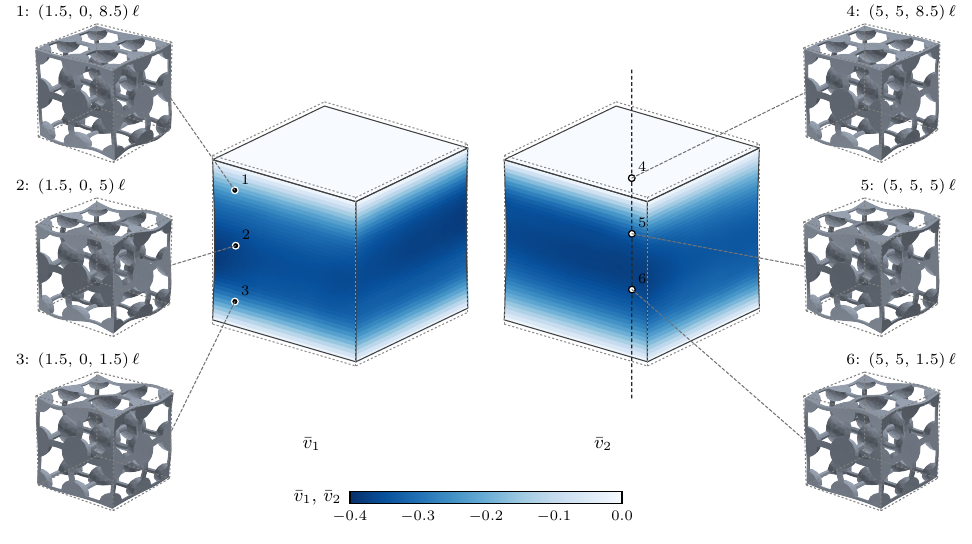}
\caption{Three-dimensional example: macroscopic solution of the MM-HROM ($N_r = M_r = 300$, $N_q = 7000$) at $u/H = 0.05$. Center: the deformed specimen (displacements to scale, undeformed outline dashed), colored by the mode amplitudes $\bar{v}_1$ (left) and $\bar{v}_2$ (right). Outer columns: the deformed RVEs (to scale, undeformed cell dashed) at the marked material points, 1--3 on the front face and 4--6 inside the specimen, on its central axis (dashed; open markers), each labeled by its reference position $\bar{\bm{X}}/\ell$, reconstructed from \cref{eq:utot} with the reduced fluctuation field.}
\label{fig:macro_3d}
\end{figure}

\begin{figure}[pos={!ht}]
\centering
\includegraphics[width=0.5\linewidth]{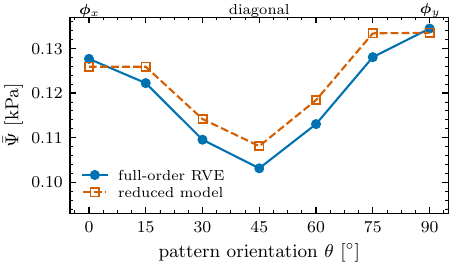}
\caption{Three-dimensional example: effective energy $\bar{\Psi}$ of the RVE at the stretch of the specimen center at $u/H = 0.05$ ($\bar{\bm{g}}_i = \bm{0}$), for patterns $(\bar{v}_1, \bar{v}_2) = a\,(\cos\theta, \sin\theta)$ of in-plane orientation $\theta$, with the amplitude $a$ minimizing $\bar{\Psi}$ along each orientation, for the full-order RVE and the adopted reduced model ($N_r = M_r = 300$, $N_q = 7000$). The orientations $\theta = 0^\circ$ and $90^\circ$ correspond to the single-plane patterns $\bm{\phi}_x$ and $\bm{\phi}_y$, and $\theta = 45^\circ$ to the diagonal one.}
\label{fig:rve_orientation_3d}
\end{figure}

\Cref{fig:macro_3d} shows the macroscopic solution of the adopted model at $u/H = 0.05$, together with the deformed RVEs at selected material points. Both amplitude fields are activated, with comparable magnitudes, $|\bar{v}_1| \approx |\bar{v}_2|$, throughout the specimen; they are nearly uniform over the cross-section and decay towards the loaded faces, on which they are suppressed. This boundary layer spans roughly three unit cells: along the specimen axis, $\|(\bar{v}_1, \bar{v}_2)\|$ reaches $90\%$ of its central value at a distance of about $2.5\ell$ from the loaded faces. The load-aligned struts consequently deflect along an in-plane diagonal, alternating between neighboring cells, i.e., in the pattern of the raw eigenvector of \cref{fig:phi_modes_3d}(a), rather than within a single plane. In the post-buckling regime the specimen contracts laterally, its secant Poisson's ratio at mid-height reaching $-0.3$ at $u/H = 0.05$. The alternating deflection of the load-aligned struts, the counter-rotation of neighboring nodes and the negative Poisson's ratio beyond the critical strain agree with the local buckling observed by \citet{Meijer2026} for slender struts. In contrast to the present result, however, \citet{Meijer2026} observe that buckling selects a single plane, i.e., only one of the two polarizations, which they also derive from a reduced model of their lattice. To verify that the diagonal selection is a property of the micromorphic model with the present RVE, and not of the reduction or of the continuation, we evaluate the full-order RVE problem of \cref{sec:formulation:micro} at the stretch $\bar{\bm{U}}$ of the specimen center at $u/H = 0.05$, with $\bar{\bm{g}}_i = \bm{0}$, for patterns $(\bar{v}_1, \bar{v}_2) = a\,(\cos\theta, \sin\theta)$ of fixed in-plane orientation $\theta$, the amplitude $a$ minimizing $\bar{\Psi}$ along each orientation. As \cref{fig:rve_orientation_3d} shows, $\bar{\Psi}$ is lowest for the diagonal pattern, $\theta = 45^\circ$, and highest for the single-plane patterns $\bm{\phi}_x$ and $\bm{\phi}_y$, $\theta = 0^\circ$ and $90^\circ$, which exceed it by $24\%$ and $30\%$ and are moreover unstable with respect to a rotation of the pattern. The reduced model reproduces the minimum at the diagonal to within $5\%$ in energy, although it underestimates the energy barrier towards the single-plane patterns. The diagonal selection is thus a property of the micromorphic model with the present RVE; whether the lattice itself behaves likewise, or selects a single plane as observed by \citet{Meijer2026}, cannot be decided without a DNS of the present specimen, which is out of reach (\cref{sec:examples:3d:convergence}).

\textit{Remark.} To exclude that the discrepancy with \citet{Meijer2026} is a numerical artefact, we carried out a number of additional studies, not reported here in detail: RVE computations with the strut radii of \citet{Meijer2026} and with nodes considerably stiffer than the struts, approximating their rigid spheres, as well as DNS of specimens of $4\times4\times4$ and $6\times6\times6$ unit cells. In these studies, the diagonal configuration was generally found as well. We therefore believe that the difference is not of numerical origin but stems from a difference in the configuration. One possible reason is the treatment of the specimen boundary: our specimens are obtained by tessellating unit cells, so that the spheres on the boundary are cut off and the boundary conditions are applied over the full loaded faces, whereas in the setup of \citet{Meijer2026} the load is presumably transmitted through the spheres only.

It should be emphasized that what the two studies above establish is convergence of the reduced model towards its own full-order limit, i.e., towards the micromorphic FE\textsuperscript{2} solution on the same RVE discretization, and not accuracy with respect to a fully resolved simulation of the specimen. Two further error sources are therefore not contained in the numbers quoted above. The first is the microscopic discretization error of \cref{sec:examples:3d:convergence}, which is of comparable magnitude. The second is the error of the micromorphic homogenization itself --- the quantity that \cref{sec:examples:2d} was able to isolate by comparing against the DNS --- which remains unquantified here and which, judging from the two-dimensional example and from the studies in~\citep{Sperling2024}, we expect to dominate both. The comparison with \citet{Meijer2026} above is qualitative only and cannot replace such an assessment.

\subsubsection{Computational cost}\label{sec:examples:3d:cost}

In contrast to \cref{sec:examples:2d:cost}, no online speed-up with respect to a full micromorphic model can be measured for this example, since no such reference could be computed here; any factor we quoted for it would rest on an extrapolation rather than on a measurement. What can be measured is the gain of the hyperreduction over the fully integrated reduced model at the same basis size, and the way the cost of the reduced model itself grows with $N_r$; both are reported in \cref{tab:timings_3d}. All runs use 27 MPI processes on a single node of the machine of \cref{sec:examples}. Since the adaptive load stepping is free to choose different increment sequences for different models, the table also lists the number of accepted increments and of macroscopic Newton iterations each model needed for the same load path.

The models need 15 to 17 increments and 58 to 86 Newton iterations in total to reach $u/H = 0.05$; the fully integrated models with $N_r \geq 300$ need two increments more than most others, since their continuation rejects a step at the bifurcation and perturbs along the critical eigenmode, whereas most other models continue directly onto the buckled branch. To compare the cost independently of the path, the table therefore reports the time per iteration over the first five increments, which are identical for all models. Under full integration it grows from $94$~s at $N_r = 200$ to $215$~s at $N_r = 500$. Choosing a larger basis than necessary is therefore expensive: the step from the adopted $N_r = 300$ to $N_r = 500$ increases the run time from $5.5$ to $9.5$~h, in order to remove a deviation of $3.7\%$ (\cref{tab:best_approx_3d}), of which the intermediate $N_r = 400$ already removes all but $1.0\%$ at $6.8$~h.

At the adopted basis size, the reduced integration rules cut the time from $5.5$~h to $2.1$, $3.6$ and $4.0$~h at $N_q = 3000$, $7000$ and $10\,000$, i.e., by factors of $2.6$, $1.5$ and $1.4$, part of which is owed to the two increments saved at the bifurcation. Per Newton iteration, the three rules give gains of $2.2$, $1.4$ and $1.3$.

The gain through hyperreduction is considerably more modest than the factor of $12$ by which the number of integration points is reduced at $N_q = 3000$. The same ceiling is visible in two dimensions, where the hyperreduction contributed a comparable factor of $1.9$ on top of the reduced basis (MM-ROM against MM-HROM in \cref{tab:timings_2d}).

Compared with the full RVE discretization, the full-order model evaluates the constitutive law at $36\,864$ integration points and solves the constrained sparse system of \cref{eq:newton} with $57\,819$ displacement unknowns, plus one back-substitution per macroscopic input for the tangent; the reduced model with $N_r = 300$ and $N_q = 7000$ evaluates the constitutive law at roughly one fifth of the integration points and solves a dense system of size $300$, the constraints of \cref{eq:c1,eq:c2,eq:c3} having been eliminated by construction. The number of unknowns is thus reduced by a factor of about 190 and the number of constitutive-law evaluations by a factor of about 5. On this basis we expect online speed-ups on the order of $10$ to $100$ with respect to the full micromorphic FE\textsuperscript{2} model, consistent with the factors of $46$ and $85$ measured in \cref{sec:examples:2d:cost}, which is what turns a computation we could not complete into one that runs in hours on a single compute node.

The offline stage is correspondingly more substantial than in two dimensions, but remains a one-off cost. On 27 cores of a node of the same machine it took $8.3$~h in total: $4.6$~h for the 256 training solves of \cref{sec:examples:3d:rom}, $1.4$~min for the POD of the 3664 snapshots, and $3.6$~h for the greedy point selection. The offline stage is therefore about twice as expensive as a single online run of the adopted model. 

What is not reported here in detail is the savings in terms of memory: the full-integration model must store the constitutive law at $36\,864$ points, whereas the hyperreduced model only needs to store it at $3000$ or $7000$ points. For history-dependent constitutive laws, this can be a decisive factor in whether a simulation fits into memory or not.

\begin{table}[pos={!ht}]
\centering
\caption{Three-dimensional example at $h_{\mathrm{M}} \approx 3.3\ell$: measured wall-clock times to reach the end of the reported loading, $u/H = 0.05$ (interpolated between the two increments bracketing it), on 27 MPI processes of a single node, for the full-integration POD study and for the three ECM rules at $N_r = 300$, together with the accepted increments and Newton iterations needed to reach it; the time per iteration is taken over the first five increments, which are identical for all models. Online speed-ups are with respect to the full-integration $N_r = 300$ run.}
\label{tab:timings_3d}
\begin{tabular*}{\tblwidth}{@{}L@{\extracolsep{\fill}}LLLLLLLL@{}}
\toprule
 & \multicolumn{5}{c}{full integration ($N_q = 36\,864$)} & \multicolumn{3}{c}{ECM at $N_r = 300$} \\
\cmidrule(lr){2-6}\cmidrule(lr){7-9}
 & $N_r = 100$ & 200 & 300 & 400 & 500 & $N_q = 3000$ & 7000 & 10\,000 \\
\midrule
wall time to $u/H = 0.05$ [h] & 2.8 & 2.4 & 5.5 & 6.8 & 9.5 & 2.1 & 3.6 & 4.0 \\
accepted load increments & 16 & 15 & 17 & 17 & 17 & 15 & 15 & 15 \\
Newton iterations & 75 & 59 & 86 & 82 & 81 & 58 & 60 & 61 \\
time per iteration [s] & 56 & 94 & 148 & 170 & 215 & 69 & 103 & 118 \\
online speed-up, wall time & --- & --- & 1.0 & --- & --- & 2.6 & 1.5 & 1.4 \\
online speed-up, per iteration & --- & --- & 1.0 & --- & --- & 2.2 & 1.4 & 1.3 \\
\bottomrule
\end{tabular*}
\end{table}

\section{Conclusions}\label{sec:conclusions}

The simulation of mechanical metamaterials is extremely challenging due to their multi-scale nature and the presence of bifurcations, making direct numerical simulations of engineering-scale specimens typically intractable. Micromorphic computational homogenization removes the need to resolve the full object and replaces it by a macroscopic continuum enriched with patterning-amplitude fields, at the price of an approximation whose accuracy is governed by the enrichment ansatz. Nevertheless, the resulting model is computationally expensive: a full-order RVE problem, with its constraints and its sensitivity solves for the tangent, must still be solved at every macroscopic integration point, in every macroscopic Newton iteration, and in every load step.

In this work, we therefore proposed a reduced order model for the microscopic problem arising in micromorphic computational homogenization, combining POD of the fluctuation field with a hyperreduction method tailored to the micromorphic setting. In both numerical examples, the reduction was assessed in two stages: first under full integration, so as to isolate the error of the POD--Galerkin projection, and subsequently with the reduced integration rule.

In two dimensions, where both a DNS and the full micromorphic model remain affordable, the micromorphic implementation was first validated against the DNS, and the reduced model then against the full micromorphic solution. With $150$ displacement modes and $8.4\%$ of the original integration points, the MM-HROM reproduces the full micromorphic response, including the local patterning and the subsequent global buckling of the specimen, with an RMS error of about $0.4$~kPa, i.e., slightly above $1\%$ of the peak stress, at a measured online speed-up of $85$ with respect to the full micromorphic model and against an offline cost of under four minutes; the residual discrepancy with respect to the DNS is attributable to the homogenization scheme rather than to the reduction. In three dimensions, where neither reference could be computed, convergence was instead established internally, in the number of retained modes and in the number of retained integration points: $N_r = 300$ modes reproduce the most resolved basis ($N_r = 500$) to within $3.7\%$ in RMS, and $N_q = 7000$ points, roughly one fifth of the full quadrature, reproduce the corresponding fully integrated model to within $0.6\%$. The three-dimensional example illustrates what we consider the main practical value of the approach --- it makes a two-scale buckling analysis of a three-dimensional metamaterial feasible in a regime in which neither the DNS nor the full micromorphic model is.

Several directions remain open. The most immediate one is the extension to geometrically parameterized microstructures. Since the reduced basis and the integration rule are built once in an offline stage, a ROM that additionally depends on a set of shape parameters would allow the buckling response of an entire family of shapes to be explored at the cost of a single training campaign, which would in turn make shape optimization and inverse design of three-dimensional buckling metamaterials realistic.

A second direction is suggested by the cost that remains. Despite the reduction achieved, the two-scale computations reported here are still expensive: the three-dimensional runs take hours on a compute node, and the expected online speed-ups of one to two orders of magnitude, while substantial, still leave the method well outside the range in which a two-scale simulation could be called cheap. The reason is structural rather than incidental: a reduced RVE problem is still solved, and its tangent still assembled, at every macroscopic integration point. Within the ROM framework, a natural remedy exploits the fact that buckling changes the microscopic fluctuation subspace significantly: instead of a single global basis spanning all configurations, adaptive subspaces could be employed, switched for each RVE according to the configuration, i.e., the buckling mode it currently resides in, so that considerably smaller bases may suffice. Another promising avenue is to employ the present reduced model as an efficient data generator for training machine-learning surrogates of the effective response; this would eliminate the microscopic problem from the online stage altogether.

\section*{Declaration of competing interest}
The authors declare that they have no known competing financial interests or personal relationships that could have appeared to influence the work reported in this paper.

\section*{Acknowledgements}
This research did not receive any specific grant from funding agencies in the public, commercial, or not-for-profit sectors.

\section*{Data availability}
The \texttt{fe2\_rom} library implementing the methods of \cref{sec:formulation,sec:rom} is openly available at \url{https://github.com/theronguo/fe2_rom}. The scripts reproducing the numerical examples of \cref{sec:examples}, together with the generated reduced bases and integration rules, can be requested from the authors.

\appendix

\section{Complete macroscopic linearization}\label{app:linearization}

For completeness, the second variation of \cref{eq:macro_energy}, of which \cref{eq:macro_lin} states the $\bar{\bm{u}}$-row only, is given here in full. Writing the increment as $(\Delta\bar{\bm{u}}, \{\Delta \bar{v}_j\})$ and the test functions as $(\delta\bar{\bm{u}}, \{\delta \bar{v}_i\})$, and abbreviating $\Delta\bar{\bm{F}} \coloneqq (\nabla\Delta\bar{\bm{u}})^T$, it reads
\begin{align}
D^2\mathcal{E}&\cdot\big[(\Delta\bar{\bm{u}}, \{\Delta \bar{v}_j\}), (\delta\bar{\bm{u}}, \{\delta \bar{v}_i\})\big] = \nonumber\\
&\quad \int_{\mathcal{B}} (\nabla\delta\bar{\bm{u}})^T : \frac{\partial\bar{\bm{P}}}{\partial\bar{\bm{F}}} : \Delta\bar{\bm{F}}\, \mathrm{d}\bar{\bm{X}}
 + \sum_{j=1}^{N_\phi} \int_{\mathcal{B}} (\nabla\delta\bar{\bm{u}})^T : \Big( \frac{\partial\bar{\bm{P}}}{\partial \bar{v}_j} \Delta \bar{v}_j + \frac{\partial\bar{\bm{P}}}{\partial\bar{\bm{g}}_j}\cdot\nabla\Delta \bar{v}_j \Big) \mathrm{d}\bar{\bm{X}} \nonumber\\
&\quad + \sum_{i=1}^{N_\phi} \int_{\mathcal{B}} \delta \bar{v}_i\, \frac{\partial\bar{\Gamma}_i}{\partial\bar{\bm{F}}} : \Delta\bar{\bm{F}}\, \mathrm{d}\bar{\bm{X}}
 + \sum_{i,j=1}^{N_\phi} \int_{\mathcal{B}} \delta \bar{v}_i \Big( \frac{\partial\bar{\Gamma}_i}{\partial \bar{v}_j} \Delta \bar{v}_j + \frac{\partial\bar{\Gamma}_i}{\partial\bar{\bm{g}}_j}\cdot\nabla\Delta \bar{v}_j \Big) \mathrm{d}\bar{\bm{X}} \nonumber\\
&\quad + \sum_{i=1}^{N_\phi} \int_{\mathcal{B}} \nabla\delta \bar{v}_i \cdot \frac{\partial\bar{\bm{\Lambda}}_i}{\partial\bar{\bm{F}}} : \Delta\bar{\bm{F}}\, \mathrm{d}\bar{\bm{X}}
 + \sum_{i,j=1}^{N_\phi} \int_{\mathcal{B}} \nabla\delta \bar{v}_i \cdot \Big( \frac{\partial\bar{\bm{\Lambda}}_i}{\partial \bar{v}_j} \Delta \bar{v}_j + \frac{\partial\bar{\bm{\Lambda}}_i}{\partial\bar{\bm{g}}_j}\cdot\nabla\Delta \bar{v}_j \Big) \mathrm{d}\bar{\bm{X}}.
\label{eq:macro_lin_full}
\end{align}
The nine distinct derivative blocks appearing here,
\begin{equation}
\frac{\partial(\bar{\bm{P}}, \bar{\Gamma}_i, \bar{\bm{\Lambda}}_i)}{\partial(\bar{\bm{F}}, \bar{v}_j, \bar{\bm{g}}_j)},
\label{eq:tangent_grid}
\end{equation}
constitute the $3\times3$ block structure of the macroscopic tangent, i.e., of the Hessian of $\bar{\Psi}$, referred to in \cref{sec:formulation:macro}. Each is obtained from the microscopic sensitivity problem of \cref{sec:formulation:effective}: the derivatives with respect to $\bar{v}_j$ and $\bar{\bm{g}}_j$ follow directly from \cref{eq:dPbar,eq:dGamma,eq:dLambda} with $\mu \in \{\bar{v}_j\} \cup \{\bar{g}_{j,k}\}$, while those with respect to $\bar{\bm{F}}$ additionally pass through the lab-frame reconstruction of \cref{eq:reconstruct_scalar,eq:reconstruct_Pbar}. By the symmetry of second derivatives of $\bar{\Psi}$, this Hessian is symmetric --- $\partial\bar{\bm{P}}/\partial \bar{v}_j$ is the transpose counterpart of $\partial\bar{\Gamma}_j/\partial\bar{\bm{F}}$, and likewise for the remaining off-diagonal pairs --- so that only six of the nine blocks need be computed independently, and the assembled macroscopic tangent is symmetric.

\printcredits

\bibliographystyle{cas-model2-names}

\bibliography{cas-refs}



\end{document}